\documentclass[prl,aps,twocolumn,superscriptaddress]{revtex4-1}
\usepackage{bm}
\usepackage{graphicx}
\usepackage{amsmath}
\usepackage{amssymb}
\usepackage[utf8]{inputenc}
\usepackage[T1]{fontenc}
\usepackage{color}
\usepackage{upgreek}
\usepackage{subfigure}
\usepackage[unicode=true,colorlinks=true,citecolor=blue,urlcolor=black]{hyperref}
\usepackage{placeins}
\usepackage{lipsum}
\usepackage{epsfig}
\usepackage{makecell}
\usepackage[utf8]{inputenc}
\usepackage{wrapfig}
\usepackage[exponent-product=\cdot]{siunitx}
\usepackage{braket}

\usepackage[normalem]{ulem}

\newcommand{\nix}[1]{}
\usepackage{nicefrac} 			

\usepackage{tikz}

\DeclareRobustCommand\circled[1]{\tikz[baseline=(char.base)]{\node[shape=circle,draw,inner sep=.6pt] (char) {\scriptsize #1};}}

\usepackage{mhchem}
\usepackage{siunitx}

\usepackage{chngcntr}

\newcommand{\J}{\mathop{\rm J}\nolimits}
\newcommand{\I}{\mathop{\rm I}\nolimits}

\begin{document}

\title{
Highly Superlinear Giant 
Terahertz Photoconductance in GaAs \\
Quantum Point Contacts in the Deep Tunneling Regime}

\author{M. Otteneder}
\affiliation{Terahertz Center, University of Regensburg, 93040 Regensburg, Germany}
    
\author{M. Hild}
\affiliation{Terahertz Center, University of Regensburg, 93040 Regensburg, Germany}
    
\author{Z. D. Kvon}
\affiliation{Rzhanov Institute of Semiconductor Physics, 630090 Novosibirsk, Russia}
\affiliation{Novosibirsk State University, 630090 Novosibirsk, Russia}
   
\author{E. E. Rodyakina} 
\affiliation{Rzhanov Institute of Semiconductor Physics, 630090 Novosibirsk, Russia}  
\affiliation{Novosibirsk State University, 630090 Novosibirsk, Russia}
   
\author{M. M. Glazov}
\affiliation{Ioffe Institute, 194021 St. Petersburg, Russia}
\affiliation{National Research University, Higher School of Economics, 190121 Saint Petersburg, Russia}

\author{S.\,D. Ganichev}
\affiliation{Terahertz Center, University of Regensburg, 93040 Regensburg, Germany}
\affiliation{CENTERA, Institute of High Pressure Physics PAS, 01142 Warsaw, Poland}

\begin{abstract}
A highly superlinear in radiation intensity photoconductance induced by terahertz laser radiation with moderate intensities has been observed in quantum point contacts made of GaAs quantum wells operating in the deep tunneling regime. For very low values of the normalized dark conductance $G_{\rm dark}/ G_0 \approx 10^{-6}$, with the conductance quantum  $G_0=2e^2/h$,  the photoconductance scales exponentially with the radiation intensity, so that already at 100 mW/cm$^2$ it increases by almost four orders of magnitude. This effect is observed for a radiation electric field oriented along the source drain direction. We provide model considerations of the effect and attribute it to the variation of the tunneling barrier height by the radiation field made possible by local diffraction effects. We also demonstrate that cyclotron resonance due to an external magnetic field manifests itself in the photoconductance completely suppressing the photoresponse.
\end{abstract}

\maketitle

\section{Introduction}
\label{introduction}

The quantum point contact (QPC) -- a narrow constriction between two  electrically-conducting regions of two-dimensional electron gas (2DEG) --  is one of the key quantum devices of physics of low-dimensional electronic systems since more than 30 years~\cite{Wees1988,Wharam1988,Glazman1988,Buettiker1990,Buettiker1992,Reznikov1995,Houten1996,Imry1997}. 
Electronic, transport, and optoelectronic properties of QPCs are the subject of extensive theoretical and experimental investigations, because these devices, on the one hand, provide an excellent model system for benchtop studies of tunneling effects and, on the other, have the potential for numerous practical applications. 
While various phenomena observable in the open state ($G \geq 2e^2/h$) of the QPC, in which the conductance quantization in units of $2e^2/h$ is detected, or in the pinch-off regime ($G \leq 2e^2/h$) are in focus of the current research, surprisingly enough, until recently the deep tunneling regime with $G \ll 2e^2/h$ was much less studied.

Most recently it has been observed that excitation of a QPC with terahertz (THz) or microwave radiation results in an increase of the QPC conductance by about 50 times if it operates in the tunneling regime with $G_{\rm dark}/ G_0 \approx 10^{-3}$, see Refs.~\cite{Levin2015,Otteneder2018}. 
These results were discussed in terms of photon-assisted tunneling. 
In the present work we demonstrate that further reduction of the dark conductance to $G_{\rm dark}/ G_0 \approx 10^{-6}$ results in a drastic enhancement of the tunneling current under irradiation, so that the conductance increases by almost four orders of magnitude. 
Moreover, studying the intensity dependence of the photoconductance we observed that under these conditions the signal exponentially grows with the radiation intensity. 
Detailed investigation of the polarization dependence of the observed effect demonstrated that both, large photoconductance magnitudes and the exponential intensity dependence, are present for the radiation electric field oriented parallel to the source-drain direction only.

We discuss microscopic mechanisms of the giant photoconductance. 
The analysis shows that the standard approach with photon-assisted tunnneling is not sufficient to describe the experimental data. 
Despite the fact that for $G_{\rm dark}/ G_0 \approx 10^{-6}$ the $I$-$V$ characteristic of the QPC is highly nonlinear and multi-photon processes are needed, the calculated photoconductivity is smaller than in the experiment. 
Our analysis demonstrates the role of the local-field effects due to the diffraction at the gates, which result in a modification of the tunneling barrier by the radiation electric field. 
The model accounting for the radiation-induced variation of the barrier height in the adiabatic approximation yields qualitative agreement with the experiment. 
For quantitative agreement one also has to take into account the local enhancement of the THz field and, possibly, an interplay of the barrier height reduction with multi-photon assisted processes.

Furthermore, we observe that the THz photoconductance can be quenched by a magnetic field oriented normally to the 2DEG plane. 
We demonstrate that the suppression of the THz-induced tunneling is most pronounced under conditions of cyclotron resonance: Sweeping the magnetic field we detected a resonant dip in the photoconductance, where the cyclotron frequency of electrons in the magnetic field corresponds to the frequency of the THz radiation.

\section{Samples and methods}
\label{samples_methods}

\begin{figure}
\centering
\includegraphics[width=\linewidth]{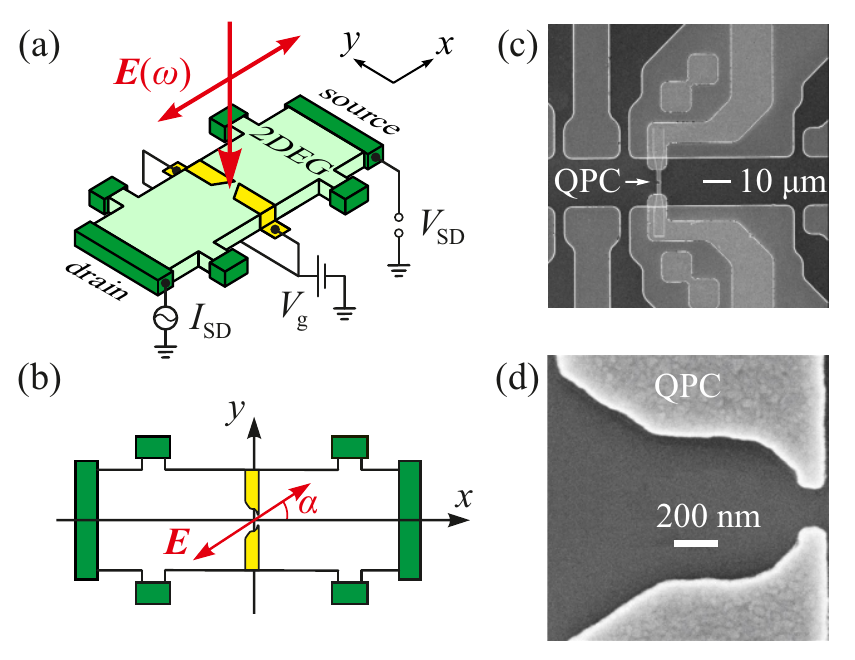}
\caption{(a) Sketch of the Hall bar shaped sample hosting the split gate QPC structure, and the measurement scheme used in the experiments. (b) Top view of the sample. (c) Microphotograph of the conduction channel and the gate fingers. (d) High magnification microphotograph of the split gate structure.}
\label{fig1}
\end{figure}

The samples investigated in this work were fabricated from a high-mobility modulation doped GaAs/(Al,Ga)As 
quantum well (QW) with a width of $\SI{12.5}{\nano\meter}$.
Material characterization by 4-terminal electrical transport measurements carried out at $T=\SI{4.2}{\kelvin}$ yield a sheet density $n_\text{s}\approx \SI[per-mode=symbol]{9e11}{\centi\meter^{-2}}$ and mobility $\mu\approx \SI[per-mode=symbol]{1.1e6}{\centi\meter\squared\per\volt\per\second}$ of the two-dimensional electrons.
This corresponds to a mean free path of around $l_\text{t}\approx \SI{20}{\micro\meter}$. 
The material was prepared into Hall bar shaped samples and on top a $\SI{120}{\nano\meter}$ thick layer of insulator and nano-meter sized split gate structures have been fabricated using electron-beam lithography.
Microphotographs of the Hall bar conduction channel and the split gate structure are presented in Figs.~\ref{fig1} (c) and (d), respectively.
%

\begin{figure}
\centering
\includegraphics[width=\linewidth]{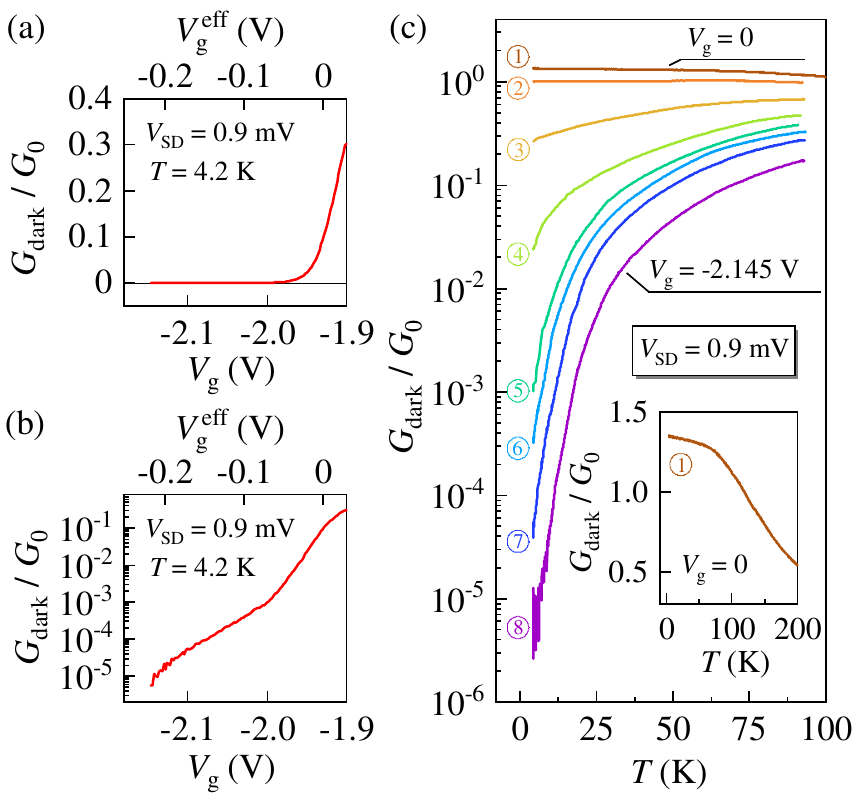}
\caption{Typical gate voltage dependence of the dark conductance $G_\text{dark}$ normalized to the conductance quantum $G_0=2e^2/h$ obtained at a temperature of $\SI{4.2}{\kelvin}$ displayed in double linear (a) and semi-logarithmic scale (b). 
This curve as well as all other subsequently presented curves in this work were obtained with an applied source-drain voltage of $\SI{0.9}{\milli\volt}$.
The upper $x$-axis shows the effective gate voltage $V_\text{g}^\text{eff}$, which was calculated according to Eq. \eqref{Vgeff}.
Panel (c) presents temperature dependencies of the dark conductance $G_\text{dark}/G_0$ obtained at different applied gate voltages (curve \circled{1} : $V_\text{g}=0$, curve \circled{2} : $V_\text{g}=\SI{-1789}{\milli\volt}$, curve \circled{3} : $V_\text{g}=\SI{-1887}{\milli\volt}$, curve  \circled{4} : $V_\text{g}=\SI{-1930}{\milli\volt}$, curve \circled{5} : $V_\text{g}=\SI{-1948}{\milli\volt}$, curve \circled{6 }: $V_\text{g}=\SI{-1978}{\milli\volt}$, curve \circled{7} : $V_\text{g}=\SI{-1991}{\milli\volt}$, curve \circled{8} : $V_\text{g}=\SI{-1994}{\milli\volt}$). 
The inset displays curve \circled{1} without applied gate voltage for a wider temperature range.}
\label{fig2}
\end{figure}

%
%
Figure~\ref{fig2} shows the results of transport measurements carried out in a 2-terminal source-drain scheme according to Fig.~\ref{fig1} (a). 
For that, an alternating voltage $V_\text{SD}$ with frequency ranging between 3 and $\SI{12}{\hertz}$ was applied to the sample and the resulting current $I_\text{SD}$ was measured with conventional low frequency lock-in amplifier technique.
Figures~\ref{fig2}~(a) and (b) show an example of a typical gate voltage dependence of the dark conductance $G_\text{dark}/G_0$ obtained at $T=\SI{4.2}{\kelvin}$ plotted in double linear and semi-logarithmic presentation, respectively.
Here, equal gate voltages 
have been applied to both parts of the split gate using RC filters, which improve stability against electrostatic discharges.
An increase of the negative gate voltage amplitude yields a decrease of the dark conductance $G_0$ by five orders of magnitude, see Fig. \ref{fig2}.
This drastic reduction demonstrates that in this range of gate voltages the deep tunneling regime is realized.
Note that the gate voltage dependencies may shift for different cooldowns.
Therefore, the upper $x$-axes in Figs. \ref{fig2}~(a) and (b) additionally show the effective gate voltage $V_\text{g}^\text{eff}$ which was calculated according to 
\begin{align}
V_\text{g}^\text{eff}=V_\text{g}-V_\text{g}^{\ast}
\label{Vgeff}
\end{align}
where $V_\text{g}^\ast$ is the gate voltage at which the dark conductance is equal to $0.1~G_0$.

Figure \ref{fig2} (c) shows the temperature dependencies of the dark conductance measured at different gate voltages corresponding to tunneling and open regime of the QPC.
These data demonstrate that in the tunneling regime ($V_\text{g}\lesssim \SI{-1.9}{\volt}$), an increase of the temperature leads to a drastic increase of the dark conductance $G_\text{dark}$.
This behaviour is in accordance with thermal tunneling of the charge carriers through the QPC potential barrier.
Without applied gate voltage, in contrast, the conductance decreases with the temperature increase, which corresponds to the temperature dependence of the GaAs two-dimensional electron gas, see the inset in Fig. \ref{fig2}~(c).

%
%
The response of the QPCs to THz illumination has been measured by irradiating the samples with linearly polarized, normally incident radiation as depicted in the setup sketched in Fig. \ref{fig1} (a).
Photoconductance $\Delta G$ was obtained as the difference between the conductance $G_\text{ph}$ measured with continuous terahertz illumination of the sample and the dark conductance $G_\text{dark}$ without incident radiation.
The samples were placed in an optical helium bath/exchange gas cryostat with $z$-cut crystalline quartz windows to couple in the terahertz radiation. 
All windows were additionally covered by black polyethylene foil, which is transparent in the THz frequency range but prevents uncontrolled illumination of the sample by visible or room light.
For photoexcitation we use a continuous wave (cw) line-tunable molecular gas laser \cite{Kvon2008,Dantscher2017}.
With methanol (\ce{CH3OH}), difluoromethane (\ce{CH2F2}) and formic acid (\ce{CH2O2}) we obtain monochromatic radiation with frequencies (photon energies) $f=\SI{2.54}{\tera\hertz}$ ($\hbar\omega=\SI{10.5}{\milli\electronvolt}$), $\SI{1.63}{\tera\hertz}$ ($\SI{6.74}{\milli\electronvolt}$) and $\SI{0.69}{\tera\hertz}$ ($\SI{2.85}{\milli\electronvolt}$), and a maximum power of around $P= \SI{50}{\milli\watt}$ , $\SI{45}{\milli\watt}$, and $\SI{13}{\milli\watt}$, respectively.
The radiation has a Gaussian beam profile and is focused onto the sample
by off-axis parabolic mirrors. 
Using a pyroelectric camera, we measured the frequency-dependent spot diameters at the sample plane, which are about $\SI{2}{\milli\meter}$, $\SI{3}{\milli\meter}$ and $\SI{3.5}{\milli\meter}$.
This parameters yield incident maximum radiation intensities of around $I=\SI[per-mode=symbol]{1500}{\milli\watt\per\centi\meter\squared}$, $\SI[per-mode=symbol]{600}{\milli\watt\per\centi\meter\squared}$ and $\SI[per-mode=symbol]{130}{\milli\watt\per\centi\meter\squared}$ depending on the respective frequency.
The polarization state of the incident radiation was varied and controlled by wire-grid or polyethylene polarizers and room-temperature half-wave plates made of $x$-cut crystalline quartz placed in the optical path before the radiation is focused onto the sample.
Most of the experiments have been carried out with a radiation electric field vector $\bm{E}$ oriented either perpendicular (see Fig. \ref{fig1} (a)) or parallel to the gate fingers.
Additionally, we studied the polarization dependence of the photoresponse. 
By rotation of the half-wave plate we changed the angle $\alpha$ between $\bm{E}$ and the $x$-axis, see Fig. \ref{fig1}~(b).
At $\alpha=0$, the electric field is oriented perpendicular to the gate fingers and parallel to the
source-drain line.
One of the important aims of this study is the investigation of the intensity dependence of the photoresponse.
To vary the intensity we used a crossed-polarizer setup consisting of two successive linear polarizers.
First, the linearly polarized laser beam passes through a rotatable polarizer which results in a decrease of the radiation intensity and a rotation of the polarization state depending on the orientation of the polarizer. 
The additional second polarizer fixed at a certain orientation causes a further decrease of the radiation intensity and creates a polarization state equal to the initial one. 
With this method, we can vary the radiation intensity between maximal (obtained for a parallel orientation of both polarizers) and zero intensity (obtained for a perpendicular orientation of the two polarizers).

An important task of this work is the study of the influence of external magnetic fields on the QPC terahertz photoconductance.
In these experiments, we apply external magnetic fields up to $\SI{3}{\tesla}$ via a liquid-helium-cooled superconducting coil, which is oriented either normal ($B_z$) or parallel ($B_x$) to the sample plane.

\section{Results}
\label{results}
\subsection{Highly superlinear THz photoconductance}

\begin{figure}
\centering
\includegraphics[width=\linewidth]{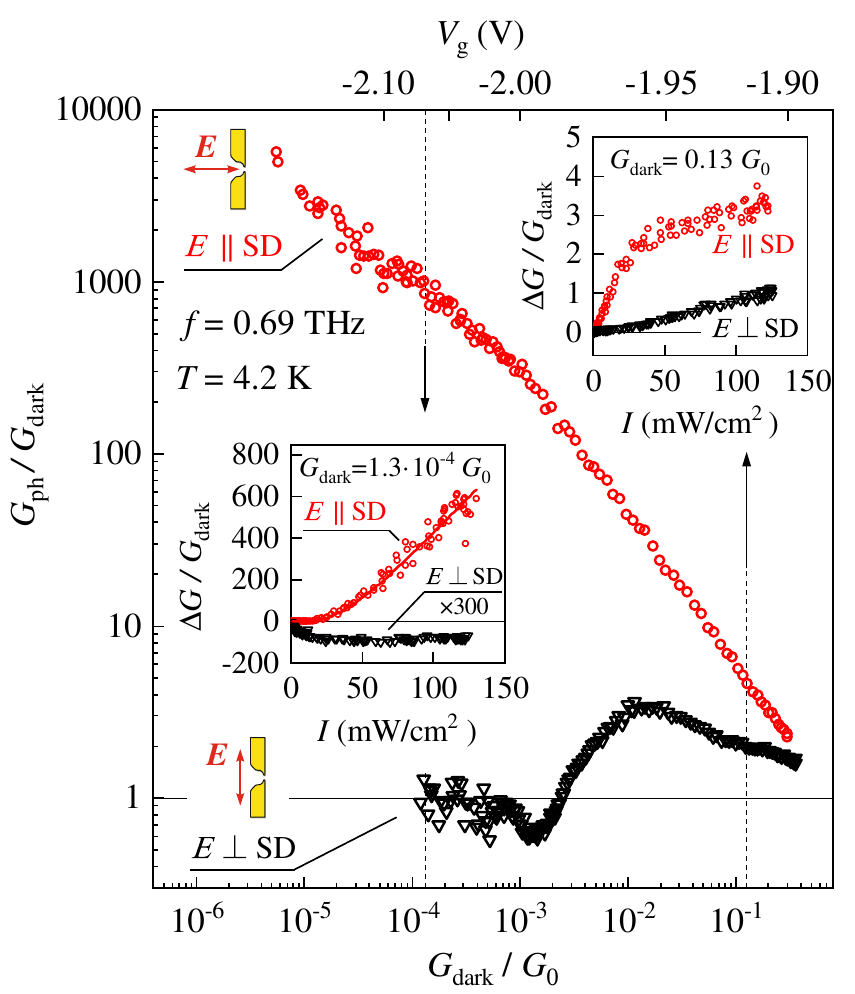}
\caption{Normalized photoresponse $G_\text{ph}/G_\text{dark}$ with respect to the dark conductance $G_\text{dark}/G_0$ presented in double logarithmic scale. 
The data are obtained at liquid helium temperature with a radiation frequency $f=\SI{0.69}{\tera\hertz}$ and intensity $I\approx\SI[per-mode=symbol]{130}{\milli\watt\per\centi\meter\squared}$ for two orthogonal orientations of the electric field vector $\bm{E}$. 
Red circles correspond to a radiation field parallel to the source-drain direction, i.e. $\bm{E}\parallel\text{SD}$, whereas black triangles correspond to $\bm{E}\perp\text{SD}$. 
The insets show the photoconductance $\Delta G /G_\text{dark}$ with respect to the radiation intensity $I$ for both electric field orientations in different dark conductance regimes. 
The data of the upper inset was obtained at a dark conductance of $G_\text{dark}=0.13~G_0$, while the bottom inset shows intensity dependencies for $G_\text{dark}=\SI{1.3e-4}{}~G_0$. 
Note that the curve for $\bm{E}\perp \text{SD}$ in the bottom inset has been multiplied by a factor of 300 for better visibility. 
The solid red line in the bottom inset presents a fit curve according to Eq. \eqref{fit1}.
}
\label{fig3}
\end{figure}

\begin{figure}
\centering
\includegraphics[width=\linewidth]{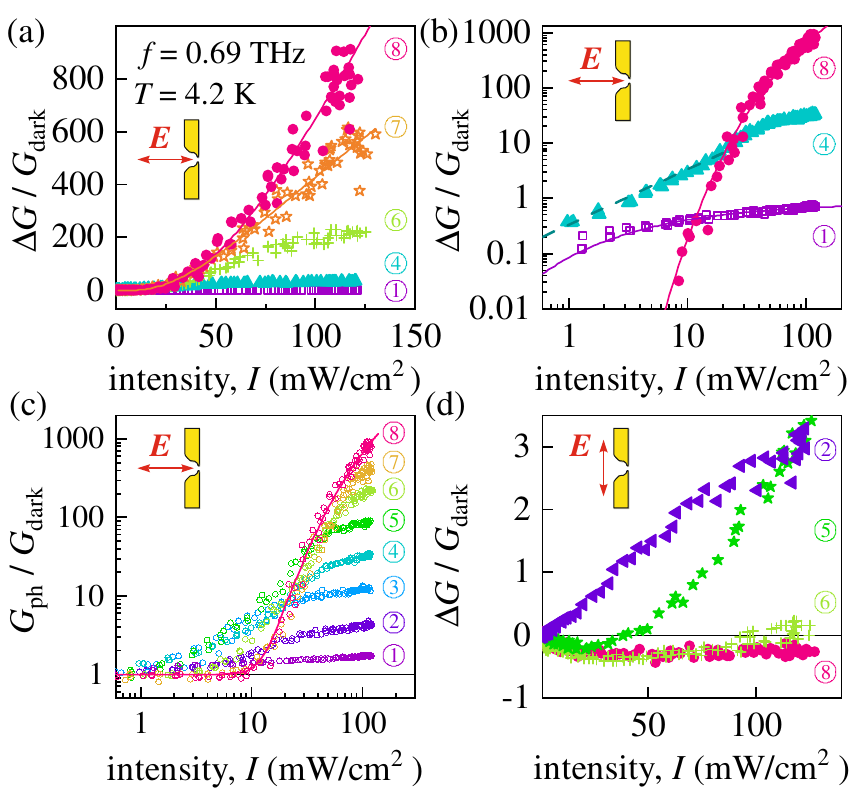}
\caption{Normalized photoconductance $\Delta G/G_\text{dark}$ with respect to the incident radiation intensity $I$ obtained for an electric field vector oriented along the source-drain direction displayed in double linear (panel (a)) and semi-logarithmic scale (panel (b)). 
Panel (c) additionally presents the photosignals $G_\text{ph}/G_\text{dark}$ obtained under the same conditions as a double logarithmic plot.
All data displayed in this figure are obtained at $T=\SI{4.2}{\kelvin}$ and $f=\SI{0.69}{\tera\hertz}$ for different values of the dark conductance $G_\text{dark}$.
In all panels curves labeled with \circled{1} correspond to $G_\text{dark}=0.43~G_0$, \circled{2} to $G_\text{dark}=0.13~G_0$, \circled{3} to $G_\text{dark}=\SI{4.5e-2}{}~G_0$, \circled{4} to $G_\text{dark}=\SI{1.3e-2}{}~G_0$, \circled{5} to $G_\text{dark}=\SI{4.3e-3}{}~G_0$, \circled{6} to $G_\text{dark}=\SI{1.4e-3}{}~G_0$, \circled{7} to $G_\text{dark}=\SI{4.7e-4}{}~G_0$ and \circled{8} to $G_\text{dark}=\SI{4.4e-5}{}~G_0$.
Note that the solid lines accompanying the data are fits following Eq. \eqref{fit1}.
The dashed dark green line in (b) is a linear curve with $\Delta G\propto I/3$.
Panel (d) shows intensity dependencies of the photoconductance $\Delta G/G_\text{dark}$ for a radiation field vector oriented normal to the SD-direction, i.e. parallel to the gate stripes.
}
\label{fig4}
\end{figure}
%
%
%
%
When illuminating the QPC samples, we observe that the incident THz radiation results in a change of the QPC conductance, which exhibits complex gate voltage and intensity dependencies.
Figure \ref{fig3} shows the dependence of the normalized photoresponse $G_\text{ph}/G_\text{dark}$ on the dark conductance $G_\text{dark}/G_0$ varied by the gate voltage applied to the gate stripes.
The data were obtained at a temperature $T=\SI{4.2}{\kelvin}$, radiation frequency $f=\SI{0.69}{\tera\hertz}$, radiation intensity $I\approx\SI[per-mode=symbol]{130}{\milli\watt\per\centi\meter\squared}$, and orientation of the electric field vector $\bm{E}$ normal and parallel to the source-drain direction (SD-direction).
Strikingly, in the deep tunneling regime with a dark conductance $G_\text{dark}\lesssim\SI{e-5}{}~G_0$, application of the THz electric field oriented along the SD-direction results in an increase of the conductance up to about four orders of magnitude.
Furthermore, the photosignal for this polarization exhibits a highly nonlinear gate voltage dependence drastically increasing for the decrease of the dark conductance.
In contrast, for radiation with $\bm{E}\perp \text{SD}$ the photoresponse decreases with the dark conductance decrease and is several orders of magnitude smaller in the deep tunneling regime than that for $\bm{E}\parallel \text{SD}$.
Note that for $G_\text{dark}\lesssim\SI{2e-3}{}~G_0$ the response drops below 1, which indicates that the incident THz radiation causes a decrease of conductance in this regime.
%
%
%
Moreover, intensity dependencies obtained for the two perpendicular orientations of $\bm{E}$ are found to be completely different in the deep tunneling regime, see bottom left inset in Fig. \ref{fig3}.
While for $\bm{E}\perp \text{SD}$ the photoconductance $\Delta G/G_\text{dark}=\left(G_\text{ph}-G_\text{dark}\right)/G_\text{dark}$ linearly decreases to negative values and saturates with rising intensity, for $\bm{E}\parallel \text{SD}$ we observe a highly superlinear dependence.
The latter can be well fitted by the empirical formula
\begin{align}
\Delta G/G_\text{dark}=A\exp{\left(-\frac{B}{E}\right)} \text{ ,}
\label{fit1}
\end{align}
as demonstrated by the solid red line in the left inset of Fig. \ref{fig3}.
In Eq. \eqref{fit1}, $A$ and $B$ are fitting parameters and $E=\sqrt{2 I Z_0/ n_\omega}$ denotes the electric field strength, with $Z_0=\sqrt{\mu_0/\epsilon_0}$, and the frequency-dependent refractive index $n_\omega$ of the medium. Here, $\epsilon_0$ and $\mu_0$ are the electric and magnetic vacuum permittivities, respectively.
In the regime of $G_\text{dark}$ approaching $G_0$, in contrast, for both configurations the intensity dependence is linear followed by saturation as shown in the right inset in Fig. \ref{fig3}.
%

%
%
Let us continue with the results of a detailed study of the intensity dependence presented in Fig. \ref{fig4}.
Figures \ref{fig4}~(a) and (b) show intensity dependencies of the photo-induced change of the conductance $\Delta G/G_\text{dark}$ obtained at different values of the dark conductance $G_\text{dark}$.
Because the intensity and the photoresponse changes by many orders of magnitude, we present the data in double linear as well as double logarithmic plots.
For the electric field vector oriented parallel to the SD-direction, we observed that Eq. \eqref{fit1} describes the data well including the increase of the amplitude by five orders of magnitude in the deep tunneling regime for $G_\text{dark}=\SI{4.4e-5}{}~G_0$, see curve \circled{8} in Fig.~\ref{fig4}~(a) and (b).
We note that with the increase of the dark conductance, the intensity dependence of the photoresponse becomes weaker (i.e. the amplitudes of the fit parameters $A$ and $B$ reduce) and eventually approaches an almost linear dependence with saturation at higher intensities at $G_\text{dark}\gtrsim \SI{e-2}{}~G_0$.
In Fig. \ref{fig4}~(c), we additionally show intensity dependencies of the photo-induced conductivity $G_\text{ph}/G_\text{dark}=\Delta G/ G_\text{dark}+1$.
This figure clearly shows the weakening of the intensity dependence with the increase of the dark conductance.
Note that in this presentation, at low intensities, all data approach unity.
\begin{figure}
\centering
\includegraphics[width=\linewidth]{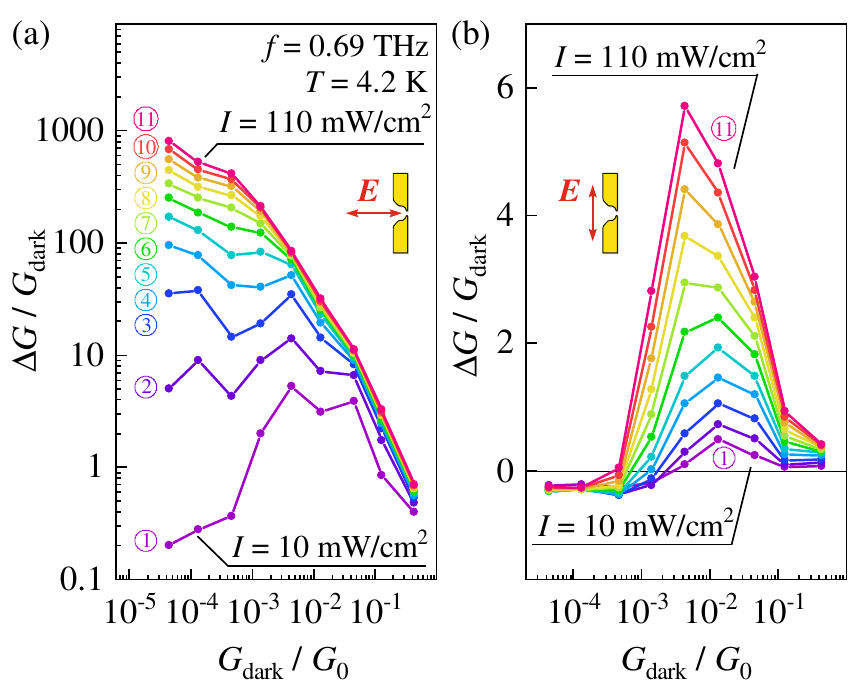}
\caption{Normalized photoconductance $\Delta G/G_\text{dark}$ as a function of the dark conductance $G_\text{dark}/G_0$ for several different incident radiation intensities $I$ ranging from $\SI[per-mode=symbol]{10}{\milli\watt\per\centi\meter\squared}$ to $\SI[per-mode=symbol]{110}{\milli\watt\per\centi\meter\squared}$.
The data points were extracted from the intensity dependencies for different dark conductance values shown in Fig. \ref{fig4}. 
Panel (a) presents data for the radiation field vector oriented parallel to the source-drain direction, i.e. $\bm{E}\parallel \text{SD}$, whereas (b) displays data obtained for $\bm{E}\perp \text{SD}$.
In panel (a) curves labeled with \circled{1} correspond to $I=\SI[per-mode=symbol]{10}{\milli\watt\per\centi\meter\squared}$, \circled{2} to $20$, \circled{3} to $30$, \circled{4} to $40$, \circled{5} to $50$, \circled{6} to $60$, \circled{7} to $70$, \circled{8} to $80$, \circled{9} to $90$, \circled{10} to $100$, and \circled{11} to $I=\SI[per-mode=symbol]{110}{\milli\watt\per\centi\meter\squared}$. 
Note that in panel (b) labeling of the curves was omitted except for curves \circled{1} and \circled{11} for better visibility. However, the color coding is identical to panel (a).
}
\label{fig5}
\end{figure}

\subsection{Polarization dependence}

While for $\bm{E}\parallel \text{SD}$, a highly nonlinear signal and drastic increase of the photo-induced conductance $\Delta G/G_\text{dark}$ are detected, for $\bm{E}\perp \text{SD}$ the photoconductance is by three orders of magnitude smaller as compared to that obtained at $\bm{E}\parallel \text{SD}$, see Fig. \ref{fig4}~(d).
For high and very low dark conductances ($G_\text{dark}\approx0.1~G_0$ and $G_\text{dark}\approx\SI{e-5}{}~G_0$, respectively), the signal amplitude first rises linearly with low intensity and then saturates, see curves \circled{2} and \circled{8} in Fig. \ref{fig4}~(d).
For the intermediate values of $G_\text{dark}$, the photoresponse exhibits a complex intensity dependence including a sign change with rising radiation intensity causing the photoconductance to be negative for low values of $G_\text{dark}$.

%
%
Studying the dependence of $\Delta G /G_\text{dark}$ on the dark conductance $G_\text{dark}$ at different radiation intensities (shown in Fig. \ref{fig5}) we observed that while for $\bm{E}\perp \text{SD}$ its behaviour remains almost the same from small up to the highest value of intensity (see Fig. \ref{fig5}~(b)), for $\bm{E}\parallel \text{SD}$ the dependence changes qualitatively, see Fig. \ref{fig5}~(a). 
At low intensities, for both orientations of $\bm{E}$, the photoresponse exhibits a non-monotonic behavior: 
With the decrease of $G_\text{dark}$ the signal increases by up to one order, approaches a maximum, and afterwards decreases, see curve \circled{1} in Fig. \ref{fig5}~(a) and all curves in Fig. \ref{fig5}~(b).
The only difference is that for $\bm{E}\perp \text{SD}$, the signal changes its sign at low dark conductance, whereas for $\bm{E}\parallel \text{SD}$ the sign inversion is not detected at the lowest possible intensity ($\SI[per-mode=symbol]{10}{\milli\watt\per\centi\meter\squared}$).
Note that the position of the sign inversion for $\bm{E}\perp\text{SD}$ shifts to smaller $G_\text{dark}$ upon the increase of the radiation intensity.
For $\bm{E}\parallel\text{SD}$ the increase of the intensity results in the qualitative modification of the dependence on dark conductance.
First of all, the increase of the radiation intensity results in a drastic (by five orders of magnitude) increase of the photoresponse at small $G_\text{dark}$, see curve \circled{11} in Fig. \ref{fig5}~(a).
As a result, instead of the non-monotonic behaviour detected at low power, we observe continuous growth of the signal upon reduction of $G_\text{dark}$, see curves \circled{6} to \circled{11} in Fig. \ref{fig5}~(a).

\begin{figure}
\centering
\includegraphics[width=\linewidth]{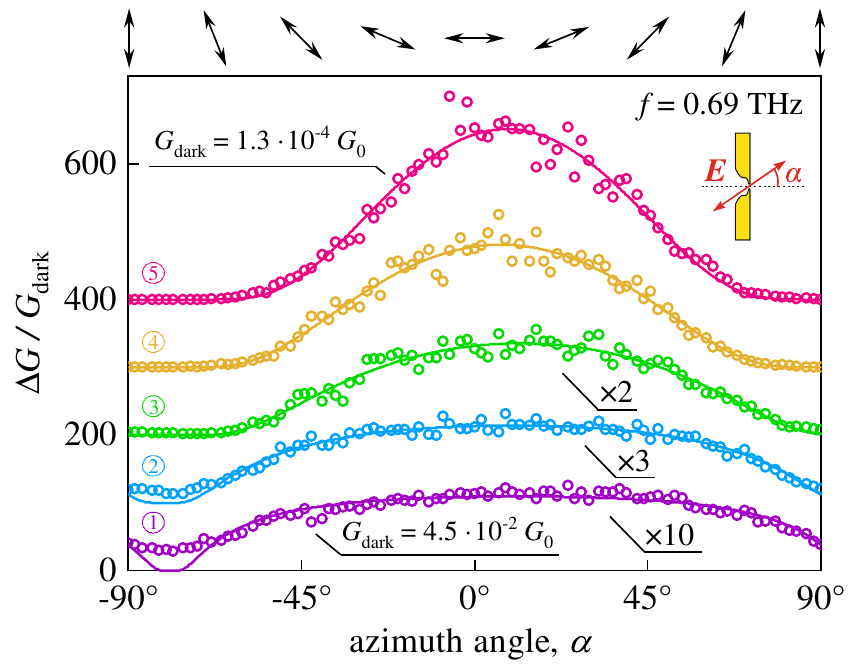}
\caption{Dependence of the normalized photoresponse $\Delta G / G_\text{dark}$ on the orientation of the electric field vector $\bm{E}$ described by the angle $\alpha$ between the SD-direction and $\bm{E}$ as defined in the inset. 
The data is obtained at $f=\SI{0.69}{\tera\hertz}$ for different conductance regimes (curve \circled{1} at $G_\text{dark}=\SI{4.5e-2}{}~G_0$, \circled{2} at $G_\text{dark}=\SI{1.2e-2}{}~G_0$, \circled{3} at $G_\text{dark}=\SI{5.2e-3}{}~G_0$, \circled{4} at $G_\text{dark}=\SI{1.2e-3}{}~G_0$ and \circled{5}  at $G_\text{dark}=\SI{1.3e-4}{}~G_0$). 
Note that the data for higher conductances have been multiplied by a constant factor of 10 (curve \circled{1}), 3 (curve \circled{2}) and 2 (curve \circled{3}). 
For better visibility the upper curves have been offset by 100 with respect to each other. 
Solid lines correspond to a fit according to equation \eqref{fit1} 
with the electric field vector $E_x=E \cos(\alpha-\alpha^\prime)$.}
\label{fig6}
\end{figure}

%
%
All previous results reveal that the photoresponse is strongly dependent on the orientation of the radiation electric field vector $\bm{E}$ with respect to the SD-direction and $G_\text{dark}$.
To explore this dependence we measured $\Delta G /G_\text{dark}$ as a function of the azimuth angle $\alpha$ for different values of dark conductance.
These data are presented in Fig. \ref{fig6}.
Note that due to a strong reduction of the signal already at moderate $G_\text{dark}$, the corresponding traces are multiplied by constant factors.
The figure reveals that at all $G_\text{dark}$ the signal approaches a maximum for the azimuth angle $\alpha\approx 0$.
Assuming that the amplitude of the signal is defined by the projection of the radiation electric field vector $\bm{E}$ on the source-drain axis $x$, we found that the polarization dependencies can be well fitted by Eq. \eqref{fit1} in which the radiation electric field vector $E$ is substituted by $E_x=E \cos(\alpha-\alpha^\prime)$. 
Here, the angle $\alpha^\prime$ accounts for a small shift ($\alpha^\prime\approx\SI{7}{\degree}$) of the maximum position away from $\alpha=\SI{0}{\degree}$.
Note that the phase shift is most probably caused by a slight asymmetry of the nm-sized QPC gate fingers or a small misalignment of the QPC structure by only few degrees with respect to the Hall bar conduction channel. 
Furthermore, we note that the best fits are obtained for small values of $G_\text{dark}$, whereas for moderate $G_\text{dark}$ a deviation is detected at $\alpha\approx\SI{90}{\degree}$, see curve \circled{1} in Fig. \ref{fig6}.

\begin{figure}
\centering
\includegraphics[width=\linewidth]{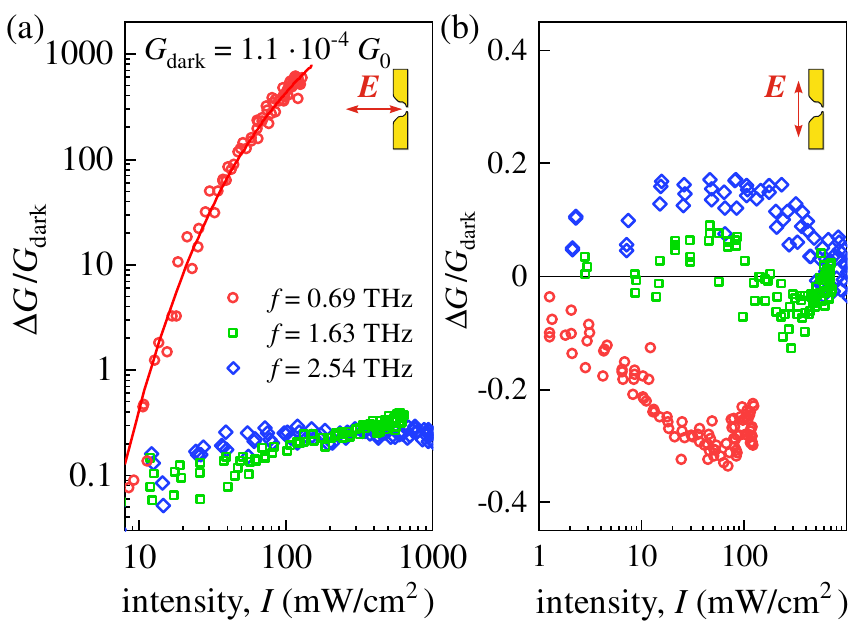}
\caption{Normalized photoconductance $\Delta G/G_\text{dark}$ with respect to incident radiation intensity $I$ shown for different radiation frequencies.
The data are obtained for a dark conductance $G_\text{dark}=\SI{1.1e-4}{}~G_0$ at $T=\SI{4.2}{\kelvin}$. Red circles correspond to $f=\SI{0.69}{\tera\hertz}$, green squares to $\SI{1.63}{\tera\hertz}$ and blue diamonds to $\SI{2.54}{\tera\hertz}$. 
Panel (a) shows data obtained for the radiation field vector parallel to SD-direction.
The solid line is a fit curve according to Eq.~\eqref{fit1}. 
In panel (b) the data for $\bm{E}$ rotated by $\SI{90}{\degree}$ is presented.
Note that panel (a) is presented in double logarithmic scaling, whereas (b) in semi-logarithmic scale.}
\label{fig7}
\end{figure}

%
%
%
Remarkably, the drastic nonlinearity and giant photoresponse have been observed for the lowest frequency only.
An increase of frequency by only about two times results in the disappearance of the highly superlinear behaviour as well as enhancement of the signal for $\bm{E}\parallel\text{SD}$, see Fig. \ref{fig7}~(a).
For $f=1.63$ and $\SI{2.54}{\tera\hertz}$ the photoresponse exhibits a sub-linear dependence on the radiation intensity and increases only by around $\SI{40}{\percent}$ at the highest intensity.
Notably under this condition a change of frequency does not substantially affect the behaviour.

Changing the orientation of the radiation electric field to $\bm{E}\perp\text{SD}$ results in a qualitative change of the dependence. 
Under these conditions, an increase of the radiation frequency causes the photoconductive response to change sign. 
Whereas for $f=\SI{0.69}{\tera\hertz}$ the photoconductance is negative (i.e. the conductance decreases due to the incident THz radiation), at $f=\SI{2.54}{\tera\hertz}$ the irradiation results in an increase of the conductance (positive photoconductance). 
In both cases the signal first increases linearly with $I$ and decreases again for higher radiation intensities. 
It is most clearly seen for the signal in response to $f=\SI{2.54}{\tera\hertz}$, which vanishes at highest intensity and, most probably, even changes the sign.
For the intermediate frequency $f=\SI{1.63}{\tera\hertz}$ the signal is almost zero and becomes very noisy, which makes it nearly impossible to properly describe its dependence on intensity.

\begin{figure}
\includegraphics[width=\linewidth]{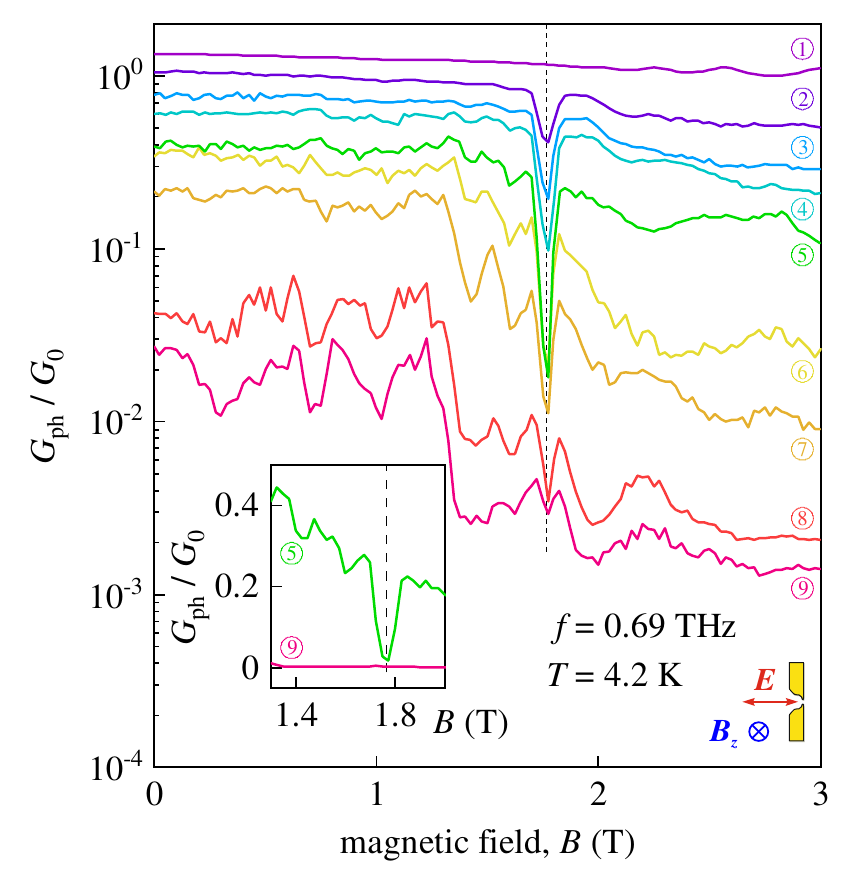}
\caption{Photoresponse $G_\text{ph}/G_0$ with respect to the out-of-plane magnetic field for an electric field oriented parallel to the source-drain direction presented in semi-logarithmic scale.
The data are obtained at $T=\SI{4.2}{\kelvin}$ and $f=\SI{0.69}{\tera\hertz}$ for different applied gate voltages corresponding to the following vales of the zero-field dark conductance $G_\text{dark}(0)$:
Curve \circled{1} corresponds to $G_\text{dark}(0)=\SI{1.35}{}~G_0$, curve \circled{2} to $G_\text{dark}(0)=\SI{9.7e-1}{}~G_0$, curve \circled{3} to $G_\text{dark}(0)=\SI{4.3e-1}{}~G_0$, curve \circled{4} to $G_\text{dark}(0)=\SI{1.7e-1}{}~G_0$, curve \circled{5} to $G_\text{dark}(0)=\SI{5.2e-3}{}~G_0$, curve \circled{6} to $G_\text{dark}(0)=\SI{2.0e-3}{}~G_0$, curve \circled{7} to $G_\text{dark}(0)=\SI{6.3e-4}{}~G_0$, curve \circled{8} to $G_\text{dark}(0)=\SI{3.9e-5}{}~G_0$, and curve \circled{9} to $G_\text{dark}(0)=\SI{4.8e-6}{}~G_0$.
Dashed lines indicate the position of the observed magneto-resonance at $B=\SI{1.76}{\tesla}$.
The inset additionally shows a zoom of curves \circled{5} and \circled{9} around the resonance in double linear presentation.
}
\label{fig8}
\end{figure}

\begin{figure}
\centering
\includegraphics[width=\linewidth]{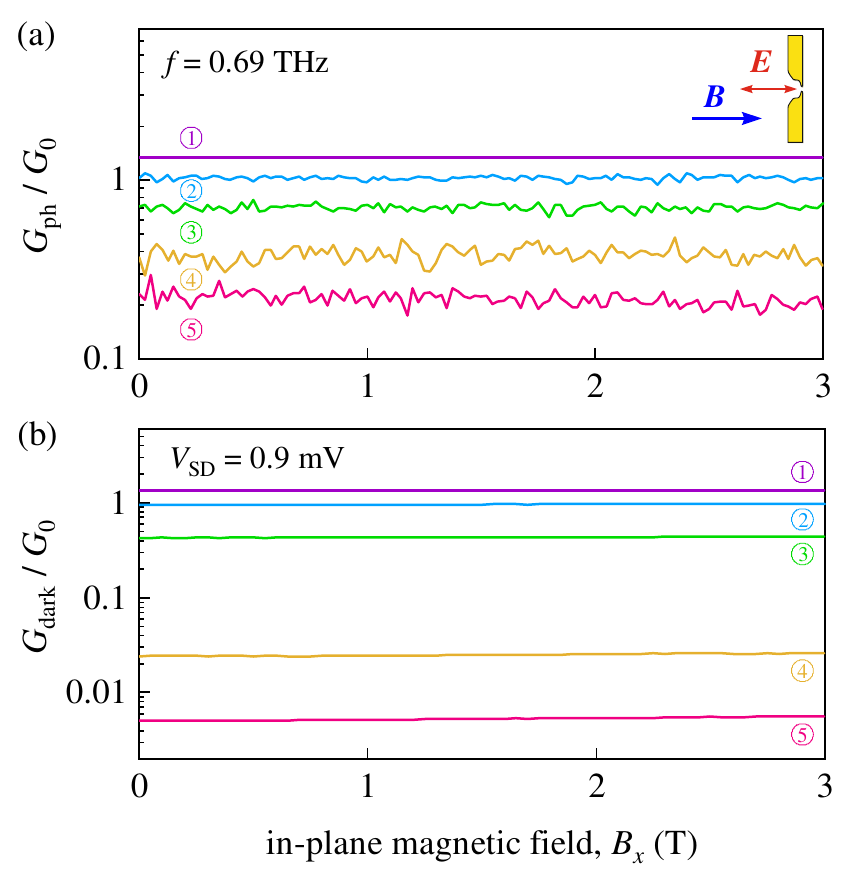}
\caption{
(a) Photosignals $G_\text{ph}/G_0$ obtained for a radiation electric field oriented along $\text{SD}$-direction as a function of the in-plane magnetic field $B_x$. 
The data are obtained at $T=\SI{4.2}{\kelvin}$ and $f=\SI{0.69}{\tera\hertz}$.
(b) Magnetic field dependence of the dark conductance $G_\text{dark}/G_0$ for a magnetic field applied in the QW plane and oriented along $\rm SD$-direction. 
The data are obtained for $V_\text{SD}=\SI{0.9}{\milli\volt}$ and several gate voltages. 
In both panels curves labeled \circled{1} correspond to an applied gate voltage of $V_\text{g}=0$, \circled{2} to $V_\text{g}=\SI{-1793}{\milli\volt}$, \circled{3} to $V_\text{g}=\SI{-1921}{\milli\volt}$, \circled{4} to $V_\text{g}=\SI{-1934}{\milli\volt}$, and \circled{5} to $V_\text{g}=\SI{-2017}{\milli\volt}$.
}
\label{fig9}
\end{figure}

\subsection{Magnetic field effects}

%
%
The most exciting result presented above -- the giant highly nonlinear photoconductance in the deep tunneling regime -- was obtained for $f=\SI{0.69}{\tera\hertz}$ and the electric field vector oriented along SD-direction.
Strikingly, this effect quenches by application of a moderate external magnetic field ($B>1$~T) oriented normal to the quantum well plane. Figure \ref{fig8} shows the magnetic field dependence of the photosignal $G_\text{ph}/G_0$ in semi-logarithmic presentation. 
Note that here we used normalization of the signal on $G_0$ instead of $G_{\rm dark}$ used above. 
This is due to a drastic increase of the dark conductivity for magnetic fields below 1~T. These results, together with the data presented in the form $G_{\rm ph}/ G_{\rm dark}$ and  $\Delta G/ G_{\rm dark}$ are addressed in Appendix~\ref{appendix}. 
Figure~\ref{fig8} demonstrates that while the photosignal is almost independent of the applied magnetic field in the range of small fields $B\lesssim\SI{1}{\tesla}$, for higher fields $B\gtrsim\SI{1.2}{\tesla}$ it causes a significant reduction of $G_\text{ph}$.
This effect becomes especially pronounced in the deep tunneling regime, where the photoresponse decreases by more than one order of magnitude at $B\gtrsim\SI{1.2}{\tesla}$, see curves \circled{6} to \circled{9} in Fig.~\ref{fig8}.
Moreover, we observe a sharp narrow dip of $G_\text{ph}$ at $B=\SI{1.76}{\tesla}$.  This resonant suppression of the photosignal is most pronounced in the region of intermediate zero-field dark conductance values $G_\text{dark}(0)\approx \SI{e-2}{}~G_0$,  see curves \circled{3} to \circled{6}  in Fig.~\ref{fig8}.
Comparison of the traces for the lowest ($G_\text{dark}(0)=\SI{4.8e-6}{}~G_0$, trace \circled{9}) and moderate ($G_\text{dark}(0)=\SI{5.2e-3}{}~G_0$, trace \circled{5}) dark conductance demonstrates that in the former case the dip is less pronounced because under these conditions the photoresponse already almost vanishes, because of the non-resonant effect of the magnetic field addressed above. 
This is demonstrated in the inset of Fig.~\ref{fig8}, which presents a zoom of the data in the vicinity of the resonance.
Here, the data are plotted in double linear presentation.
Furthermore, Fig.~\ref{fig8} shows that the depth of the resonant dip has a non-monotonic dependence on the dark conductance: with the conductance decrease it first increases, achieves a maximum at $G_{\rm dark}/G_0 \approx \SI{5e-3}{}$ and decreases to zero for the lowest values of the dark conductance.

Changing the orientation of the external magnetic field relative to the radiation electric field vector $\bm{E}\parallel \text{SD}$, we observe that both the magnetic field induced suppression of the signal as well as the magnetoresonance disappear for $\bm{B}\parallel \bm{E}\parallel\text{SD}$, see Fig. \ref{fig9}~(a). 
Note that, in contrast to the results obtained for $\bm B$ oriented normal to the QW plane (see Fig.~\ref{fig_app2}), the in-plane magnetic field does not change the dark conductance, see Fig.~\ref{fig9}~(b).

\begin{figure}
\centering
\includegraphics[width=\linewidth]{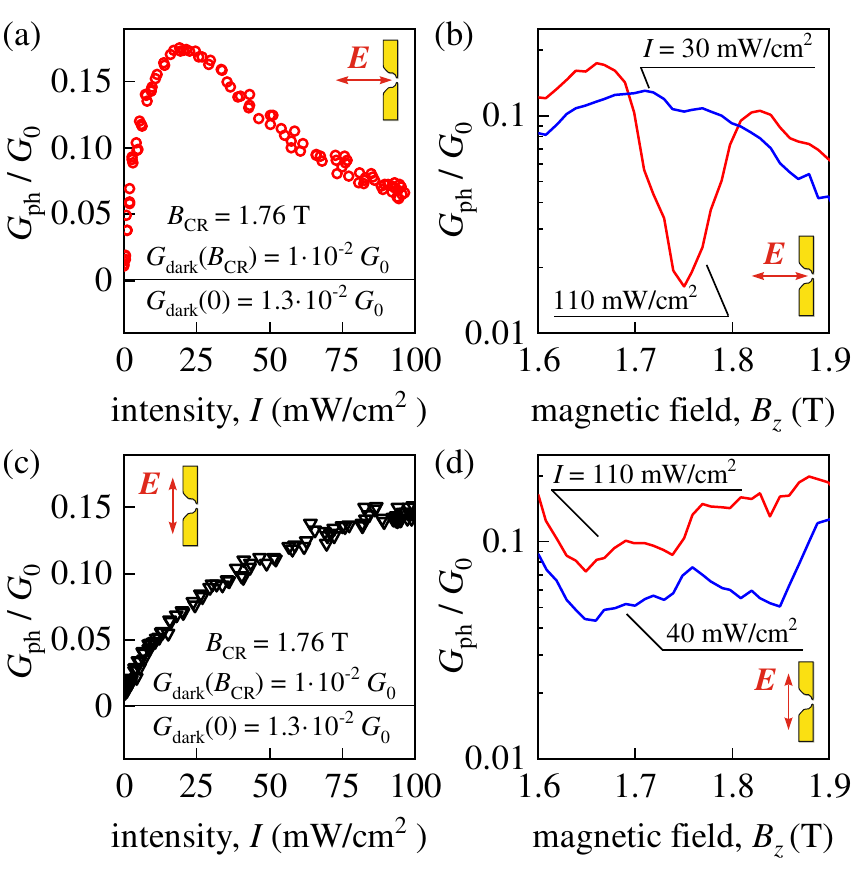}
\caption{Panels (a) and (c): Photosignal $G_\text{ph}/G_0$ as a function of the radiation intensity $I$. The data are obtained  for $B_z=B_\text{CR}$ and a zero field dark conductance of $G_\text{dark}(0)=\SI{1.3e-2}{}~G_0$, which at $B=\SI{1.76}{\tesla}$ changes to $G_\text{dark}(B=\SI{1.76}{\tesla})=\SI{1e-2}{}~G_0$, see Appendix~\ref{appendix}.
Panel (a) shows the results for the radiation field vector $\bm E \parallel {\rm SD}$, and panel (c) for $\bm E \perp {\rm SD}$ as sketched in the insets.
Panels (b) and (d): Zooms of the $B_z$ dependence of the photosignal $G_\text{ph}/G_0$ around the resonance at $B_\text{CR}$ for a radiation electric field vector oriented parallel to the source-drain direction (b) and normal to SD-direction (d), respectively.
The data were obtained at a zero-field dark conductance of $G_\text{dark}(0)=\SI{1.3e-2}{}~G_0$ for two radiation intensities $I=30$ and $\SI[per-mode=symbol]{110}{\milli\watt\per\centi\meter\squared}$ in panel (b) and $I=40$ and $\SI[per-mode=symbol]{110}{\milli\watt\per\centi\meter\squared}$ in panel (d), respectively.
Note that all data presented in this figure are acquired at $T=\SI{4.2}{\kelvin}$ and $f=\SI{0.69}{\tera\hertz}$.
}
\label{fig10}
\end{figure}

%
%
Finally we address the intensity dependence of the photoresponse under resonance condition.
All results on the magneto-photoresponse discussed so far were obtained for the highest radiation intensity.
Varying the radiation intensity, we observed that at the resonance, the intensity dependence of the photoresponse changes qualitatively as shown in Fig. \ref{fig10}~(a). 
At low intensities, the signal $G_\text{ph}/G_0$, alike at zero magnetic field, increases nonlinearly with rising intensity.
At higher intensities, however, an increase of the radiation intensity results in a substantial reduction of the photosignal.
Comparing the magnetic field dependencies in the vicinity of the resonance at $B=1.76$ T for high and low intensities we observed that the resonant dip is almost absent at low intensities, see Fig. \ref{fig10}~(b).
At last but not least, we emphasize that in contrast to the configuration with $\bm{E}\parallel\text{SD}$, for the radiation electric field vector oriented normally to the source-drain-direction the resonant dip is not observed in the whole range of studied intensities, see Fig.~\ref{fig10}~(d).
Furthermore, the intensity dependence itself remains almost the same as for zero magnetic field, see Fig. \ref{fig10}~(c).

\section{Discussion}
\label{discussion}

An increase of the QPC conductance due to excitation with low power terahertz or microwave radiation has been detected previously in similar structures \cite{Levin2015,Otteneder2018,Tkachenko2021}. 
The effect is usually attributed to photon-assisted tunneling induced by the incident electromagnetic field, where the electron absorbs one or several photons during its passage through the tunneling structure~\cite{Tien1963, Tucker1979}. 
Within the simplest possible approach the effect of the radiation is included as a periodic variation of the potential of one of the leads, e.g., of the right lead in the QPC as
\[
U_\text{R} = e V_{ac} \cos{\omega t} \text{ ,}
\] 
where $e$ is the electron charge, $V_{ac}$ is the amplitude of the potential modulation and $\omega$ is the radiation frequency. 
As a result one obtains for the \emph{dc} photocurrent \cite{Tien1963, Tucker1979,Stolz2014,Ward2010,Platero2004}
\begin{equation}
\label{TG} 
I_{\rm ph} = \sum_{n=-\infty}^\infty \J_n^2\left(\frac{eV_{ac}}{\hbar\omega} \right) I_{\rm SD}\left(V_{\rm SD}+ \frac{n\hbar\omega}{e}\right) \text{ .}
\end{equation}
Here $I_{\rm SD}(V_{\rm SD})$ is the $I$-$V$ characteristic of the QPC in the absence of irradiation and $\J_n(z)$ is the Bessel function of the first kind.

For relatively low and moderate radiation intensities and not too low dark conductance the function $I_{\rm SD}(V_{\rm SD})$ is not steep, and one can retain only the terms with $n=0,\pm 1$ in Eq.~\eqref{TG} resulting in $G_{\rm ph}\propto |V_{ac}|^2 \propto I$~\cite{Stolz2014}. 
For instance, the data in Ref.~\cite{Otteneder2018} were obtained for $G_\text{dark}\gtrsim \SI{e-3}{}~G_0$ and rather low radiation intensities around $I\approx \SI[per-mode=symbol]{50}{\milli\watt\per\centi\meter\squared}$.
It was shown that, under these conditions, photon-assisted tunneling describes the results well.
Under such conditions the photoconductive signal $\Delta G/G_\text{dark}$ indeed scales linearly with radiation intensity, as illustrated by the dashed line in Fig.~\ref{fig4}~(b). 

\begin{figure}
\centering
\includegraphics[width=\linewidth]{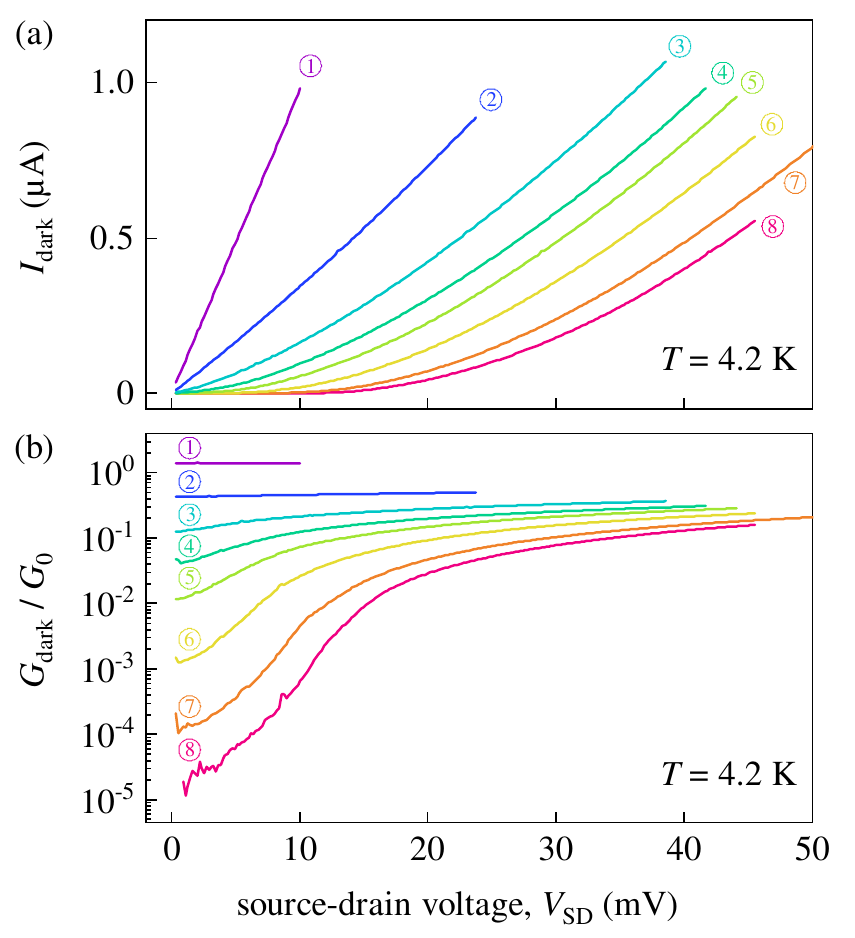}
\caption{
Characteristic dependence of the dark current $I_\text{dark}$ (a) and normalized dark conductance $G_\text{dark}/G_0$ (b) on the applied source-drain voltage $V_\text{SD}$. 
The curves are measured at different applied gate voltages corresponding to different values of dark conductance $G_\text{dark}$. 
In both panels curves labeled with \circled{1} are obtained at $V_\text{g}=0$, \circled{2} at $V_\text{g}=\SI{-1990}{\milli\volt}$, \circled{3} at $V_\text{g}=\SI{-2047}{\milli\volt}$, \circled{4} at $V_\text{g}=\SI{-2067}{\milli\volt}$, \circled{5} at $V_\text{g}=\SI{-2084}{\milli\volt}$, \circled{6} at $V_\text{g}=\SI{-2105}{\milli\volt}$, \circled{7} at $V_\text{g}=\SI{-2127}{\milli\volt}$, and \circled{8} at $V_\text{g}=\SI{-2140}{\milli\volt}$.
All data are obtained at $T=\SI{4.2}{\kelvin}$.
}
\label{fig11}
\end{figure}

The situation changes strongly in the deep tunneling regime, where the $I$-$V$ dependence becomes very steep. 
Figure \ref{fig11}~(a) shows that in deep tunneling regime, the variation of the source-drain voltage results in the exponential growth of the current.
The data for different gate voltages can be well fitted by the empirical formula
\begin{align}
\label{I-V:exper}
I_\text{SD}=G(V_\text{g}) V_\text{SD}\exp\left(\frac{-B(V_\text{g})}{|V_\text{SD}|}\right),
\end{align}
with $G>0$ and $B>0$ being gate voltage-dependent coefficients. 
Due to the exponential factor in Eq.~\eqref{I-V:exper} the electron transmission through the QPC is virtually impossible for small bias and becomes significant at $V_{\rm SD} \gtrsim B$.
In this nonlinear regime the conductance reaches $G(V_{\rm g}) \simeq G_0$, see Fig.~\ref{fig11}~(b). 
With the increase of the gate voltage the dependence becomes weaker and for $V_\text{g}=0$ the current scales linearly with $V_\text{SD}$.

For sufficiently large $B(V_{\rm g})$, such that the dimensionless parameter
\begin{equation}
\label{beta}
\beta = \frac{|eB(V_{\rm g})|}{\hbar\omega} \gg 1 \text{ ,}
\end{equation}
the photocurrent through the QPC appears only with assistance of several photons resulting in rather sharp variation of the photoconductance with the radiation intensity. 
The parameter $\beta$ gives an estimate of the number of photons needed to overcome the tunneling barrier and provide high conductance in the system. 
Indeed, in the studied experimental situation for the case of small dark conductance, $G_{\rm dark} \lesssim \SI{e-4}{}~G_0$, the parameter $B \simeq 20\ldots 30$~mV and the number of photons needed to establish high conductance is $\beta \simeq 10$.
 
\begin{figure}
\includegraphics[width=\linewidth]{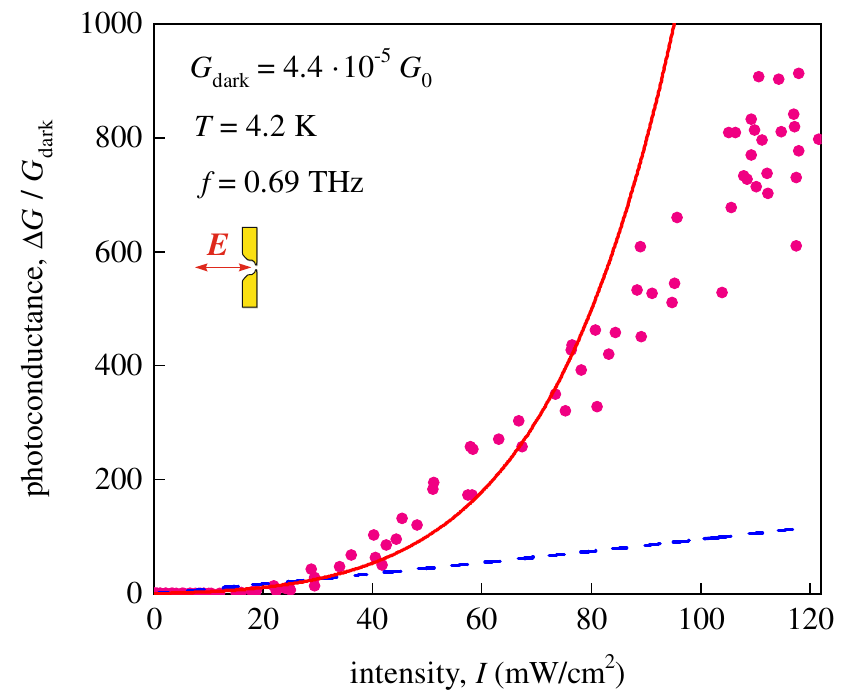}
\caption{Comparison of the experimental and theoretical results. 
Magenta circles show the experimental data for $G_\text{dark}=\SI{4.4e-5}{}~G_0$ (trace \circled{8} from Fig.~\ref{fig4}).
The blue curve shows the photoconductivity response calculated after Eq.~\eqref{G:photo:TG} for $\beta=5$ as a function of $I\propto v_{ac}^2$. 
Here, the proportionality coefficient was chosen such that at $v_{ac}=1$ the intensity equals $I=\SI[per-mode=symbol]{120}{\milli\watt\per\centi\meter\squared}$, see text for details. 
The solid red line displays the photoconductivity calculated in the adiabatic approach after Eq.~\eqref{Gph:adiab}, assuming that at $I=\SI[per-mode=symbol]{120}{\milli\watt\per\centi\meter\squared}$ the argument of the Bessel function in Eq.~\eqref{Gph:adiab}, 
equals $z=\varkappa_0 a {|e\tilde V_{ac}|}/{U}=10$.  
}
\label{fig12}
\end{figure}

In the limit of small source-drain voltages $V_{\rm SD} \ll B(V_{\rm g}), \hbar\omega/e$ one obtains from Eqs.~\eqref{TG} and \eqref{I-V:exper} the following expression for the photoconductivity
\begin{equation}
\label{G:photo:TG}
G_{\rm ph} = 2G \sum_{n=1}^\infty \J_n^2(v_{ac}) e^{-\beta/n}\left(1+ \frac{\beta}{n}\right), \quad v_{ac}=\frac{eV_{ac}}{\hbar\omega},
\end{equation}
where $v_{ac}$ is the dimensionless amplitude of the radiation-induced potential modulation. 
The blue curve in Fig.~\ref{fig12} shows the result of the calculation of the normalized photoconductance as a function of the intensity of radiation. 
To calculate the blue curve we took $\beta=5$ and assumed that $I=120$~mW/cm$^2$ corresponds to $v_{ac} = 1$. 
Note that Fig.~\ref{fig12} demonstrates the photoconductance variation normalized to $G_{\rm dark}$. Since for the $I$-$V$ characteristic in the form of Eq.~\eqref{I-V:exper} the dark conductance at $V_{\rm SD} \to 0$ vanishes, we used, as a value of the dark conductance the value of $G_{\rm ph}$ at $v_{ac}=0.1$.

While the prediction of the photon-assisted tunneling model demonstrates that the effect can be substantial
\footnote{Note that for $\beta=11$ and $v_{ac}=1$ one obtains an increase of the photonductance by $\sim 10^3$.}
, the dependence of the photoconductance $\Delta G/G_\text{dark}$ on the radiation intensity is less steep than in the experiment. 
Also, at the maximum intensity $I\approx \SI[per-mode=symbol]{150}{\milli\watt\per\centi\meter\squared}$ the radiation-induced electric field is around $E \approx 4$~V/cm. 
Assuming that this field results in a voltage drop over several $100$~nm distance we obtain $v_{ac} \simeq 0.1$, yielding a weaker photoconductivity variation as compared to the experiment. 
Better agreement can be achieved by taking into account that the field in the vicinity of the QPC is enhanced by the presence of the metallic electrodes, since the field enhancement factor $\gtrsim 10$ cannot be ruled out in our structures~\cite{Klimov1994,Kotelnikov1995,Shulman1999,Ganichev2005,Olbrich2016}.  

Moreover, an important feature of our system is that, besides the direct action of the $x$-component of the radiation electric field $E_x$ on the electrons, the terahertz radiation also results in a reduction of the barrier height due to an electric field component normal to the quantum well plane $E_z$.
The latter originates from the near field effect caused by diffraction at the spikes of the metal split gate structure. 
Indeed, as discussed in Ref.~\cite{Otteneder2018}, for a radiation electric field vector aligned parallel to SD-direction, the resulting field component $E_z$ is directed along $z$-direction for one half of the oscillation period and along $-z$-direction for the other.
This, in turn, leads to a time-dependent variation of the QPC potential barrier, which is raised for one half of the period and lowered for the other.
Qualitatively, the giant photoconductance effect can be understood within the adiabatic approach where 
\begin{equation}
\label{ad}
\omega \tau_{\rm tun} \ll 1,
\end{equation}
where $\tau_{\rm tun}$ is the tunneling time. 
Assuming that the QPC can be considered as a barrier with height $U$ and width $a$ we, following Ref.~\cite{Buettiker1982}, obtain for the tunneling time (under deep tunneling condition) $\tau_{\rm tun} = a/\sqrt{2U/m^\ast}$ with $m^\ast$ being the effective electron mass. 
The condition~\eqref{ad} thus can be recast as
\begin{equation}
\label{ad:1}
\omega \tau_{\rm tun} \simeq \frac{\hbar\omega}{U} \varkappa_0 a \ll 1,
\end{equation}
where $\varkappa_0 = \sqrt{2m^\ast U/\hbar^2}$ is the electron wavevector under the barrier. 
For $U\simeq \SI{30}{\milli\electronvolt}$ and $a\simeq \SI{100}{\angstrom}$ this condition can be fulfilled in our system
\footnote{Another possible mechanism for the giant increase of the photoconductivity could be the tunneling ionization of the QPC in the electric field of the THz radiation similarly to the ionization of atoms by electromagnetic radiation~\cite{Keldysh1965}. 
The theory of Ref.~\cite{Keldysh1965} indeed predicts an exponential dependence of the ionization rate on the \emph{ac} field strength in the adiabatic regime, where the electron tunneling under the field-induced triangular barrier is fast compared to the frequency $\omega$ of the alternating field. This condition is described by the dimensionless Keldysh-parameter $\gamma =\omega \sqrt{2mU}/|eE|$, which should be smaller than unity. However, in our case this process is weak since the adiabaticity parameter for this regime $\gamma \sim 10^3$. By contrast the condition of fast tunneling through the QPC, Eq.~\eqref{ad:1}, is readily fulfilled.}.

Under condition~\eqref{ad:1} we present the tunneling current as
\begin{equation}
\label{It}
I(t)= G_0 T(t) V_{\rm SD},
\end{equation}
where $T(t)$ is the time-dependent transmission coefficient through the barrier. 
Physically, the electron traversing the barrier experiences the momentary value of the barrier height $$U+e\tilde V_{ac}(t),$$ 
where $\tilde V_{ac}(t)$ is the potential variation due to the irradiation. Correspondingly, both the transmission coefficient and tunneling current oscillate with time~\cite{Buettiker1982}. 
We are interested in the time-averaged value of the current defined as
\begin{equation}
\label{Idc}
I_{\rm ph} = \frac{\omega}{2\pi} \int_0^{2\pi/\omega} I(t) dt.
\end{equation}
To determine $I_{\rm ph}$ and the photoconductivity in the adiabatic regime we use the quasiclassical expression for $T(t)$ in the form
\begin{equation}
\label{Tt}
T(t) = T_0 \exp{\left[-2\varkappa(t) a\right]} = T_{\rm dark} \exp{\left(-\varkappa_0 a \frac{e \tilde V_{ac}(t)}{U} \right)},
\end{equation} 
 where we made use of the approximate relation 
\[
\varkappa(t) \approx \varkappa_0\left(1+ \frac{e \tilde V_{ac}(t)}{2U} \right),
\]
and $T_{\rm dark} = T_0 \exp{(-2\varkappa_0 a)}$. Equation~\eqref{Tt} is valid for weak modulation of the potential $|e \tilde V_{ac}| \ll U$, where the time-dependence in the exponent can be taken into account only. 
Assuming a harmonic modulation of the potential $\tilde V_{ac}(t) = \tilde V_{ac}\cos{\omega t}$ we obtain for the photoconductance
\begin{equation}
\label{Gph:adiab}
G_{\rm ph} = G_{\rm dark} \I_0\left(\varkappa_0 a \frac{|e\tilde V_{ac}|}{U} \right),
\end{equation}
where $\I_0(z)$ is the modified Bessel function. 
We recall that 
\begin{equation}
\I_0(z) \approx 
\begin{cases}
1+ z^2/4, \quad z\ll 1,\\
e^z/(2\pi z), \quad z \gg 1.
\end{cases}
\end{equation}
Even if $|e\tilde V_{ac}|/U\ll 1$, for a sufficiently wide barrier $z = \varkappa_0 a  |e\tilde V_{ac}|/U$ can be significant, $z\gtrsim 1$, and the photoconductance increases exponentially with the radiation intensity. The exponential increase of the conductance in the presence of the radiation can be understood from simple physical arguments:
Since the transmission coefficient depends exponentially on the barrier height its small reduction results in a huge photoresponse.
The photoconductivity calculated after Eq.~\eqref{Gph:adiab} and normalized in such a way that at $I=120$~mW/cm$^2$, the argument of the Bessel function equals $z=\varkappa_0 a {|e\tilde V_{ac}|}/{U} =10$ is presented in Fig.~\ref{fig12} by the red curve. 
Figure~\ref{fig12} shows that this model describes the behavior of the experimental data up to $I\approx 80$~mW/cm$^2$ reasonably well. 
We attribute further deviations of the theoretical curve from the experimental points to saturation effects and inapplicability of Eq.~\eqref{Gph:adiab} for sufficiently large field amplitudes. 
Regarding the particular values of the effect, $z\simeq 10$ at $I\simeq 100$~mW/cm$^2$, substantial enhancement of the field in the vicinity of the QPC is required. 
It is also possible that the combination of photon-assisted processes and variation of the barrier height should be taken into account to describe the giant photoconductance observed experimentally.

Note that experimentally the photoconductive response of several orders of magnitude is observed only for the lowest radiation frequency $f=\SI{0.69}{\tera\hertz}$, see Fig.~\ref{fig7}~(a). 
This is in line with our analysis above: The effect of the variation of the barrier height is most pronounced in the adiabatic regime where $\omega\tau_{\rm tun}$ is as low as possible. 
Still, the precise origin of the suppression of the effect for higher frequencies and its quantitative description are open questions. 
Also, possibly, processes of multi-photon absorption in QPCs, which are accompanied by ``scattering'' on the barrier and require momentum conservation, can be important and are expected to be suppressed for higher frequencies~\cite{Alperovich2021}.

\begin{figure}
\centering
\includegraphics[width=\linewidth]{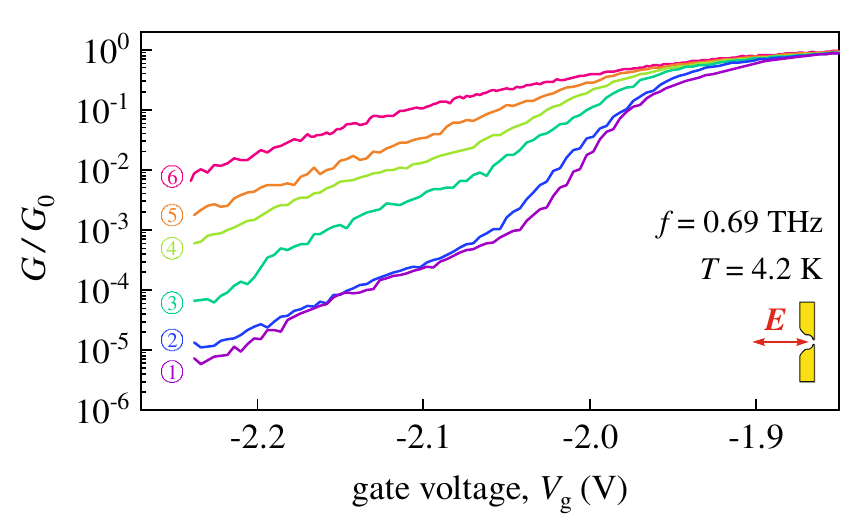}
\caption{Gate voltage dependencies of the conductance $G/G_0$ through the QPC without incident terahertz radiation ($G_\text{dark}/G_0$, curve \circled{1}) and for an incident radiation electric field $\bm{E}\parallel\text{SD}$ with $f=\SI{0.69}{\tera\hertz}$ at different intensities ($G_\text{ph}/G_0$, curves \circled{2} to \circled{6}).
Curve \circled{2} corresponds to a radiation intensity of $I=\SI[per-mode=symbol]{10}{\milli\watt\per\centi\meter\squared}$, \circled{3} to $\SI[per-mode=symbol]{30}{}$, \circled{4} to $\SI[per-mode=symbol]{50}{}$, \circled{5} to $\SI[per-mode=symbol]{70}{}$, and \circled{6} to $I=\SI[per-mode=symbol]{150}{\milli\watt\per\centi\meter\squared}$.
All data are obtained at $T=\SI{4.2}{\kelvin}$.
}
\label{fig13}
\end{figure}
%
%
%
The assumption that the effect of the THz radiation is similar to a reduction of the barrier height
is supported by measurements of the gate voltage dependencies of the QPC conductance $G_\text{ph}/G_0$ measured at different intensities of cw THz radiation including $G_\text{dark}/G_0$ at zero intensity (i.e. absent terahertz illumination) presented in Fig.~\ref{fig13}.
It is seen that incident terahertz radiation in fact leads to a horizontal shift of the $G_\text{ph}/G_0$ curves (traces \circled{2} to \circled{6} in Fig.~\ref{fig13}) into the range of higher negative gate voltages compared to the $G_\text{dark}/G_0$ curve (trace \circled{1} in Fig.~\ref{fig13}).
The higher the incident terahertz radiation intensity, the larger this shift to more negative gate voltages.
Consequently, under THz excitation, larger negative gate voltages have to be applied to obtain a similar value of conductance as without incident THz radiation.
As seen in Fig.~\ref{fig13} this shift becomes significant only for applied gate voltages corresponding to the tunneling regime of the QPC.
In this regime, it can be assumed that equal values of conductance correspond to equal barrier heights.
Thus, the observed gate voltage shift of the conductance curves under THz excitation indicates a decrease of the potential barrier due to the incident radiation field.
Assuming that the barrier height decrease is proportional to the incident terahertz radiation intensity, we obtain the exponential dependence of the QPC photoconductance on radiation intensity as observed in the experiment in the deep tunneling regime.

Importantly, as the analysis in Ref.~\cite{Otteneder2018} showed, the $E_z$ component caused by diffraction is almost homogeneous on the scale of the QPC for a incident field polarized along the source-drain direction. 
By contrast, for an incident electric field oriented normal to SD-direction, the dependence of the $E_z$-component on the spatial coordinate becomes more complex and has opposite signs for opposite sides of the gate spikes. 
Therefore, the barrier is effected by $E_z$ and $-E_z$ components simultaneously and, consequently, the radiation effect on the barrier height is expected to be weaker~\cite{Otteneder2018}. 
This is in line with the experimental results obtained for a radiation electric field oriented 
normal to the source-drain direction, i.e. parallel to the gate stripes.
Figures \ref{fig3}, \ref{fig4}~(d), \ref{fig5}~(b), \ref{fig7}~(b) show that under this condition the photoresponce in the whole range of the dark conductance is rather small so that $G_{\rm ph} \lesssim 5 G_{\rm dark}$.
Furthermore, its behaviour upon variation of $G_\text{dark}$, radiation intensity, and frequency is qualitatively different with respect to $\bm{E}$ oriented parallel to SD-direction.
For large values of $G_{\rm dark}$ the photoconductive response increases the QPC conductance and depends linearly on radiation intensity, see inset in Fig.~\ref{fig3}. 
In the deep tunneling regime and at low intensities, the sign of the photoresponse reverses, so that under these conditions irradiation results in a decrease of the conductance.
However, an increase of the radiation intensity changes the sign of photoconductance from negative to positive, see Figs.~\ref{fig4}~(d) and \ref{fig5}~(b). 
Figure~\ref{fig5}~(b) furthermore shows that the dependence of the photoresponse on $G_\text{dark}$ is qualitatively different for a radiation electric field oriented parallel to the gate stripes.
In the $\bm{E}\perp$ SD configuration, the reduction of $G_\text{dark}$ leads to a non-monotonic dependence: the photoconductance first increases, exhibits a maxiumum, decreases, and finally changes sign, see Fig.~\ref{fig5}~(b).
Here, the values of $G_\text{dark}$ at which the sign inversion takes place reduces upon increase of the radiation intensity.
Besides the difference in the intensity and $G_\text{dark}$ dependencies, we also observed that, while for $\bm{E}\parallel $ SD the incident radiation results in a drastic increase of the signal amplitude (by 4 orders of magnitude), for $\bm{E}\perp$ SD configuration, the amplitude of the signal does not change much upon variation of frequency, see Fig.~\ref{fig7}.
Here, however, we observed that the increase of frequency results in the change of sign of photoconductance.
We attribute this behaviour to electron gas heating resulting, at low intensities, in the decrease of the carriers mobility (negative $\mu$-photoconductivity caused by scattering on phonons \cite{Ganichev2005}), and, at high intensities, an increase of the tunneling probability.
For more details on this mechanism see Refs. \cite{Otteneder2018,Levin2015}. 
More detailed discussion of these effects is out of scope of the present paper focused on giant THz photoconductivity in the deep tunneling regime.

%
%
Finally, we discuss the effect of an external magnetic field on the photoconductance, in particular, the observed magnetoresonance. 
Figure~\ref{fig8} reveals that for magnetic fields $B \leq 1$~T the normalized photoresponse $G_\text{ph}/G_0$ is almost independent on the magnetic field $B_z$. At higher magnetic fields however, we observed that, already at $B\approx 1.4$ T, the photosignal $G_\text{ph}/G_0$, measured for a QPC dark conductance in the tunneling regime, reduces by one order of magnitude or even more as compared to that measured at zero magnetic field 
\footnote{Note that for $B > 1$~T the dark conductance becomes almost independent of the magnetic field, see Appendix~\ref{appendix}. Therefore, in this range of magnetic fields normalization on $G_\text{dark}(B)$ does not significantly affect the magnetic field dependence of the photoresponse, see Fig.~\ref{fig_app2}.}, see Fig.~\ref{fig8}.
The observed decrease of the photoresponse at moderate magnetic fields can be explained by the magnetic field induced suppression of the tunneling current.
Such suppression has previously been detected for both terahertz~\cite{Moskalenko1999, Ganichev2003} and static \cite{Eaves1986} electric fields \cite{Eaves1986, Moskalenko1999, Ganichev2003}.
This effect is related to the winding of the electron trajectory in the magnetic field.
Quantum-mechanically it was explained by the increase of the tunneling time for magnetic fields applied perpendicular to the electron velocity $v_x$.
The fact, that a magnetic field oriented along SD-direction ($B\parallel v_x$) does not affect the photoresponse, see Fig.~\ref{fig9}~(a), supports this mechanism. 

Apart from the monotonous reduction of the photoresponse detected at high negative gate voltages, we additionally observed a magnetoresonance, particularly pronounced at moderate values of the gate voltage, see Figs.~\ref{fig8}, \ref{fig_app1}, and \ref{fig_app3}.
Under these conditions, we detected a sharp dip in the photoresponse, whose position ($B=1.76$ T) corresponds to the cyclotron resonance of electrons in the GaAs two-dimensional system~\cite{ Warburton1992,Kono1995,Skierbiszewski1998,Heron1998,Heron2000}. 
Note that in these references other THz laser lines were used and comparison was obtained recalculating $B_{\rm CR} \propto f$. 
For CR in transmission in similar QW structures see Ref.~\cite{Herrmann2016}.
This fact reveals that the origin of the resonance is the cyclotron motion of free carriers in the two-dimensional electron gas, which localizes/traps the carriers and therefore reduces the tunneling current.

\section{Summary}
\label{summary}

In summary, our observations show that the electric field of terahertz radiation oriented along the source-drain direction drastically affects dc tunneling in quantum point contacts. 
The observed exponential dependence of the photoconductance on the radiation intensity is analyzed in the framework of two models: (i) photon-assisted tunneling and (ii) tunneling through a barrier with time-dependent height. 
The analysis shows that the effect is most likely caused by the modification of the tunnel barrier, which for one half of the period of the electromagnetic wave results in the giant enhancement of the tunneling current through the QPC. 
The strongest nonlinearity and highest amplitude of photoconductance up to 4 orders of magnitude is detected in the deep tunneling regime with the normalized dark conductance $G_{\rm dark}/ G_0 \approx 10^{-6}$ for  the highest intensities used in this work $I\approx \SI[per-mode=symbol]{130}{\milli\watt\per\centi\meter\squared}$.
An increase of $G_{\rm dark}$, obtained by a reduction of the magnitude of the negative gate voltages, greatly reduces the radiation induced change of the conductance $\Delta G$.
Furthermore, under these conditions the signal scales almost linearly with radiation intensity.

For a radiation electric field oriented perpendicular to the source-drain direction the photoconductance effect is less striking: it has significantly smaller magnitude, can be positive as well as negative, and exhibits a different dependence on dark conductivity and intensity. 
For this polarization orientation the photoconductance is attributed to THz radiation induced electron gas heating resulting in the stimulation of tunneling as well as in the change of the carrier mobility.  
Additional experiments demonstrated that the photoconductance caused by the enhancement of the tunneling current can be quenched by an external magnetic field oriented perpendicular to the source-drain direction and QW plane. 
The effect is shown to be caused by the Lorentz force acting on moving electrons and, most strikingly manifests itself in the complete quenching of the photoconductance under conditions of cyclotron resonance. 

Overall, our experiments and theoretical analysis demonstrate the importance of the local modification of the electromagnetic field of the incident THz radiation by the split gate QPC structure.
This modification gives rise to a giant, polarization-dependent, photo-induced change of the conductance by several orders of magnitude.

\section{Acknowledgments}
\label{acknow}

We thank 
D.\,A. Kozlov
for helpful discussions. The support from the 
FLAG-ERA program (project DeMeGRaS, project GA501/16-1 of the Deutsche Forschungsgemeinschaft, DFG),
the IRAP program of the Foundation for Polish Science (grant MAB/2018/9, project CENTERA), and the Volkswagen Stiftung Program is gratefully acknowledged.
Z.\,D.\,K. and E.\,E.\,R. acknowledge the support of the RFBR (Grants No. N 20-02-00385).

\appendix
\section{Magnetoresponse}
\label{appendix}
\counterwithin{figure}{section}
\setcounter{figure}{0}

\begin{figure}
\centering
\includegraphics[width=\linewidth]{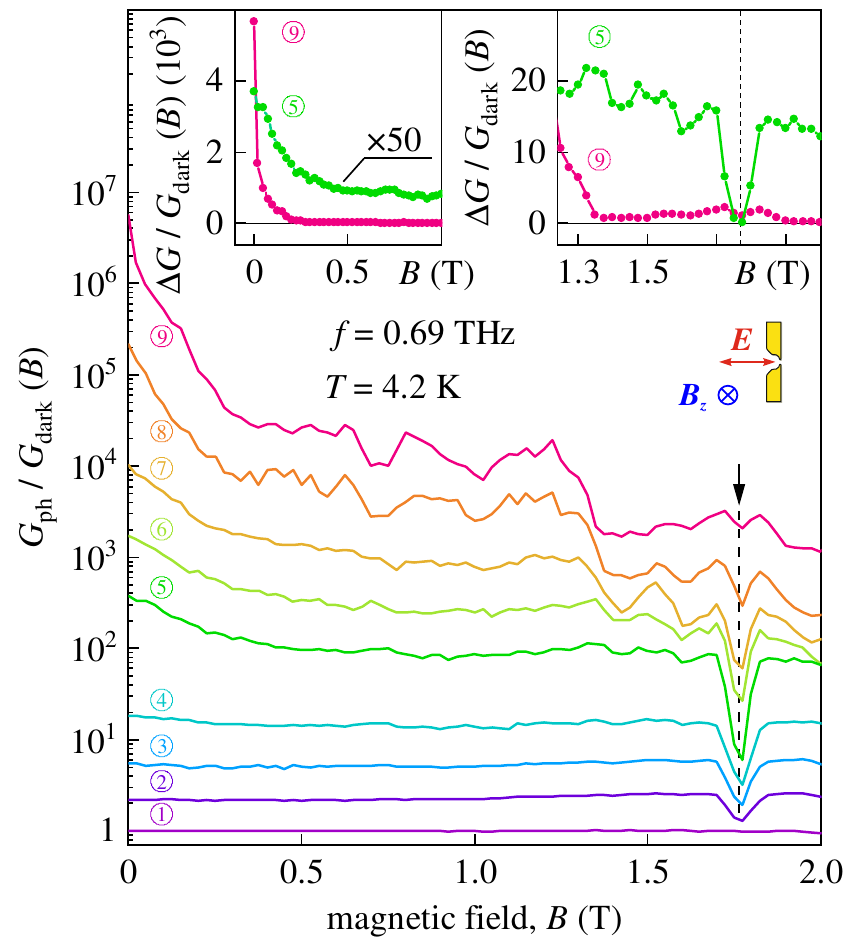}
\caption{Normalized photoresponse $G_\text{ph}/G_\text{dark}(B)$ with respect to magnetic field obtained at $T=\SI{4.2}{\kelvin}$ and $f=\SI{0.69}{\tera\hertz}$ plotted in semi-logarithmic scaling.
The curves are obtained for a radiation electric field vector $\bm{E}$ parallel to the source-drain direction at different values of the dark conductance at zero magnetic field $G_\text{dark}(0)$. 
Curve \circled{1} corresponds to $G_\text{dark}(0)=\SI{1.35}{}~G_0$, curve \circled{2} to $G_\text{dark}(0)=\SI{9.7e-1}{}~G_0$, curve \circled{3} to $G_\text{dark}(0)=\SI{4.3e-1}{}~G_0$, curve \circled{4} to $G_\text{dark}(0)=\SI{1.7e-1}{}~G_0$, curve \circled{5} to $G_\text{dark}(0)=\SI{5.2e-3}{}~G_0$, curve \circled{6} to $G_\text{dark}(0)=\SI{2.0e-3}{}~G_0$, curve \circled{7} to $G_\text{dark}(0)=\SI{6.3e-4}{}~G_0$, curve \circled{8} to $G_\text{dark}(0)=\SI{3.9e-5}{}~G_0$, and curve \circled{9} to $G_\text{dark}(0)=\SI{4.8e-6}{}~G_0$.
Note that all curves except \circled{1} are multiplied by constant factors for better visibility.
Curve \circled{2} is multiplied by 2, \circled{3} by 3, \circled{4} by 5, \circled{5} by 5, \circled{6} by 10, \circled{7} by 30, \circled{8} by 200, and \circled{9} by 1000.
The upper left inset additionally shows the low-field part of the photoconductance $\Delta G/G_\text{dark}$ for traces \circled{5} and \circled{9} in double linear presentation.
Note that curve \circled{5} is multiplied by a factor of 50 for better comparability.
In the upper right inset the high-field part of the magnetodependence of $\Delta G/G_\text{dark}$ is presented for traces \circled{5} and \circled{9}.
The dashed line in the main panel and the right inset indicates the position of the magnetoresonance at $B_\text{CR}=\SI{1.76}{\tesla}$. 
}
\label{fig_app1}
\end{figure}

\begin{figure}
\centering
\includegraphics[width=\linewidth]{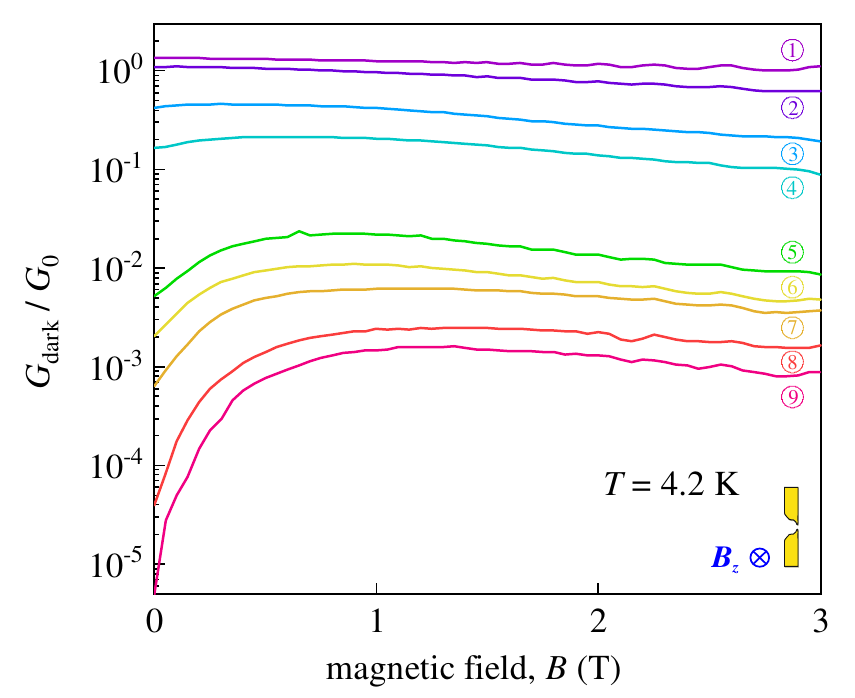}
\caption{Dark conductance $G_\text{dark}/G_0$ with respect to the out-of-plane magnetic field measured at $T=\SI{4.2}{\kelvin}$ and $V_\text{SD}=\SI{0.9}{\milli\volt}$ for different applied gate voltages.
Curve \circled{1} is obtained at $V_\text{g}=0$, \circled{2} at $V_\text{g}=\SI{-1780}{\milli\volt}$, \circled{3} at $V_\text{g}=\SI{-1900}{\milli\volt}$, \circled{4} at $V_\text{g}=\SI{-1940}{\milli\volt}$, \circled{5} at $V_\text{g}=\SI{-1987}{\milli\volt}$, \circled{6} at $V_\text{g}=\SI{-1990}{\milli\volt}$, \circled{7} at $V_\text{g}=\SI{-2019}{\milli\volt}$, \circled{8} at $V_\text{g}=\SI{-2104}{\milli\volt}$, and \circled{9} at $V_\text{g}=\SI{-2140}{\milli\volt}$.}
\label{fig_app2}
\end{figure}

\begin{figure}
\centering
\includegraphics[width=\linewidth]{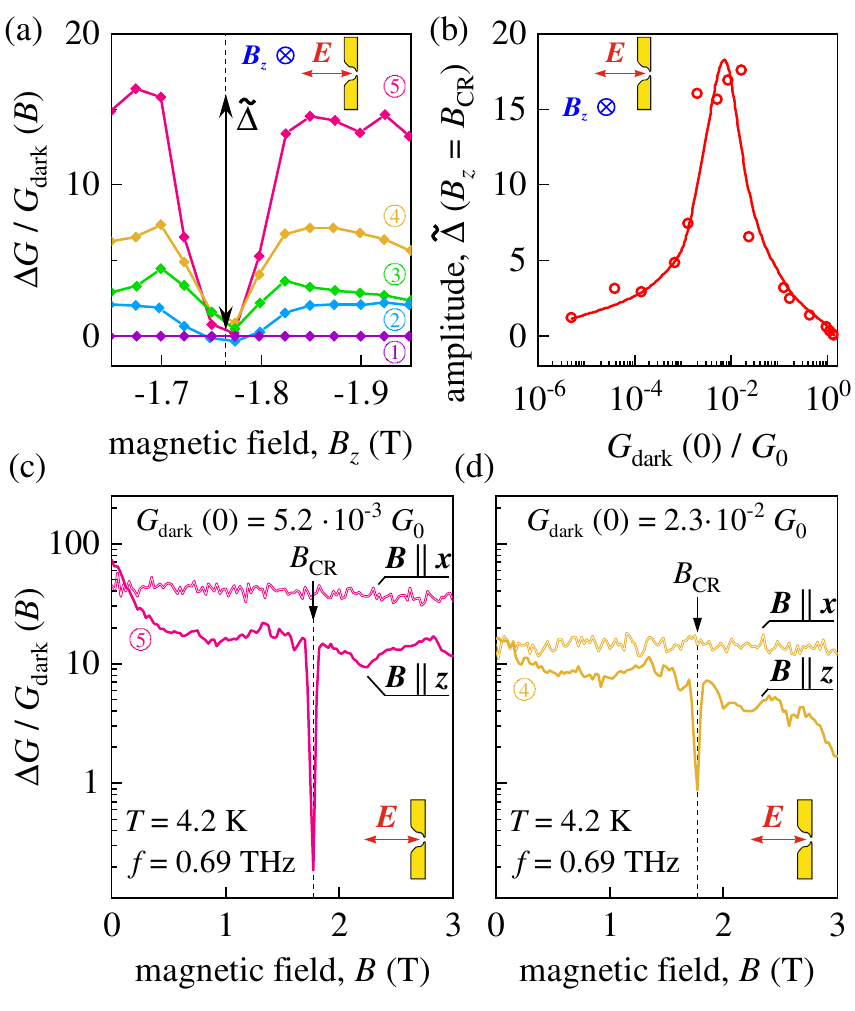}
\caption{(a) Zoom of the magnetic field dependencies of the normalized photoconductance $\Delta G/G_\text{dark}(B)$ around the resonant dip at $B_\text{CR}=\SI{1.76}{\tesla}$. 
The curves are obtained for different values of the dark conductance at zero field $G_\text{dark}(0)$, 
where trace \circled{1} corresponds to $G_\text{dark}(0)=1.3~G_0$, \circled{2} to $G_\text{dark}(0)=\SI{1.7e-1}{}~G_0$, \circled{3} to $G_\text{dark}(0)=\SI{1.3e-1}{}~G_0$, \circled{4} to $G_\text{dark}(0)=\SI{2.3e-2}{}~G_0$, and \circled{5} to $G_\text{dark}(0)=\SI{5.2e-3}{}~G_0$.
The dip amplitude $\widetilde{\Delta}$ at $B_\text{CR}=\SI{1.76}{\tesla}$ has been extracted for several curves as sketched in panel (a) and is plotted with respect to $G_\text{dark}(0)$ in panel (b).
Note that the solid line represents a guide for the eye.
Panels (c) and (d) display comparisons of the magnetic field dependencies of the photoconductance $\Delta G / G_\text{dark}(B)$ for an out-of-plane orientation of the magnetic field ($\bm{B}\parallel z$) and a field oriented parallel to the radiation electric field vector ($\bm{B}\parallel \bm{E} \parallel x$).
The data in panel (c) were obtained at a zero field dark conductance of $G_\text{dark}(0)=\SI{5.2e-3}{}~G_0$, whereas the data in panel (d) correspond to $G_\text{dark}(0)=\SI{2.3e-2}{}~G_0$.
}
\label{fig_app3}
\end{figure}

In the main section we demonstrated that, for the QPC in the tunneling regime and $\bm{B}\parallel\text{SD}$-direction, the photoconductance $G_\text{ph}$ normalized on  $G_\text{0}$ is almost independent of the magnetic field for $B<1$~T, substantially decreases for higher $B$-fields and exhibits a sharp resonant dip at $B = 1.76$~T, see Fig.~\ref{fig8}. 
Replotting these data with normalization on $G_\text{dark}(B)$ we found that with this data presentation the normalized photoconductance drastically decreases already at very small magnetic fields, see Fig.~\ref{fig_app1} curves \circled{9} to \circled{5}. 
This drastic difference in the magnetic field dependence of $G_\text{ph}/G_\text{0}$ and $G_\text{ph}/G_\text{dark}(B)$ is caused by a steep increase of the dark conductance $G_\text{dark}$ in magnetic field, shown in Fig.~\ref{fig_app2}.
Here, the magnetic field dependence of the dark conductance $G_\text{dark}/G_0$ obtained for a source drain voltage $V_\text{SD}=\SI{0.9}{\milli\volt}$ is shown for different applied gate voltages. The qualitative behaviour of $G_\text{dark}$ strongly depends on the gate voltage applied to the QPC structure.
Whereas for highly negative applied $V_\text{g}$ corresponding to the deep tunneling regime, $G_\text{dark}$ surprisingly increases considerably with the increase of the magnetic field strength for $B\leq\SI{1}{\tesla}$ (curves \circled{5} to \circled{9}), for zero or small negative values of $V_\text{g}$ (curves \circled{1} to \circled{4}) the dark conductance becomes almost independent of $B$ in the region of low magnetic fields.
Note that at the lowest values of dark conductance used in our experiments rather small magnetic fields of about 0.5~T result in the increase of $G_\text{dark}$ by about two orders of magnitude, see curve \circled{9} in Fig.~\ref{fig_app2}. 
We attract attention to the fact that for magnetic fields above 1~T the dark conductance becomes independent of magnetic field, nevertheless having different values for different gate voltages.
The origin of the increase of the dark conductance at low magnetic fields is unclear and requires further study.

Finally, for completeness, we replot the data of Figs.~\ref{fig8} and \ref{fig_app1} using normalization of the photoconductance on the dark conductance, $\Delta G/G_\text{dark}(B)$, see Fig.~\ref{fig_app3}.
Note that the resonance of the photosignal at $B=\SI{1.76}{\tesla}$ is clearly visible for all kinds of data normalization.
%
%
Figure~\ref{fig_app3}~(a) presents a zoom of the magnetic field dependence of the normalized photoconductance $\Delta G/G_\text{dark}(B)$ in the vicinity of the resonance.
At the resonant magnetic field, the photoconductive signal is suppressed by almost 20 times for $G_\text{dark}(0)=\SI{5.2e-3}{}~G_0$, see curve \circled{5} in Fig.~\ref{fig_app3}~(a).
Extraction of the dip amplitude $\widetilde{\Delta}$ as sketched in the inset of Fig.~\ref{fig_app3}~(a) shows that it has a non-monotonic dependence on the dark conductance: with the dark conductance decrease the dip amplitude first increases, achieves a maximum at $G_{\rm dark}/G_0 \approx 10^{-2}$ and decreases to zero for the lowest values of the dark conductance, see Fig.~\ref{fig_app3}~(b).
The resonant and non-resonant suppression of the photoconductance $\Delta G/G_\text{dark}$ under application of external magnetic fields is only detected for fields oriented along $z$-direction and is absent for a magnetic field orientation parallel to the radiation electric field vector ($\bm{B}\parallel \bm{E}\parallel x$), as demonstrated in Fig.~\ref{fig_app3}~(c) and (d).

\bibliography{QPC_2}
\end{document}